\documentclass[longauth]{aa}  

\usepackage{graphicx}
\usepackage{xcolor}
\usepackage{txfonts}
\usepackage{lipsum}
\usepackage[overload]{empheq}
\usepackage{amssymb}
\usepackage{pifont}
\usepackage{multirow}
\usepackage{hyperref}
\newcommand{\cmark}{\ding{51}}%
\newcommand{\xmark}{\ding{55}}%

\begin{document}

\title{A catalogue of dual-field interferometric binary calibrators}

\author{M.~Nowak\inst{\ref{cam}, \ref{kavli}}\fnmsep\thanks{Corresponding Author, mcn35@cam.ac.uk}
 \and S.~Lacour\inst{\ref{lesia},\ref{esog}}
 \and R.~Abuter\inst{\ref{esog}}
 \and A.~Amorim\inst{\ref{lisboa},\ref{centra}}
 \and R.~Asensio-Torres\inst{\ref{mpia}}
 \and W.~O.~Balmer\inst{\ref{jhupa},\ref{stsci}}
 \and M.~Benisty\inst{\ref{ipag}}
 \and J.-P.~Berger\inst{\ref{ipag}}
 \and H.~Beust\inst{\ref{ipag}}
 \and S.~Blunt\inst{\ref{northwestern}}
 \and A.~Boccaletti\inst{\ref{lesia}}
 \and M.~Bonnefoy\inst{\ref{ipag}}
 \and H.~Bonnet\inst{\ref{esog}}
 \and M.~S.~Bordoni\inst{\ref{mpe}}
 \and G.~Bourdarot\inst{\ref{mpe}}
 \and W.~Brandner\inst{\ref{mpia}}
 \and F.~Cantalloube\inst{\ref{lam}}
 \and B.~Charnay\inst{\ref{lesia}}
 \and G.~Chauvin\inst{\ref{cotedazur}}
 \and A.~Chavez\inst{\ref{northwestern}}
 \and E.~Choquet\inst{\ref{lam}}
 \and V.~Christiaens\inst{\ref{liege}}
 \and Y.~Cl\'enet\inst{\ref{lesia}}
 \and V.~Coud\'e~du~Foresto\inst{\ref{lesia}}
 \and A.~Cridland\inst{\ref{leiden}}
 \and R.~Davies\inst{\ref{mpe}}
 \and R.~Dembet\inst{\ref{lesia}}
 \and J.~Dexter\inst{\ref{boulder}}
 \and A.~Drescher\inst{\ref{mpe}}
 \and G.~Duvert\inst{\ref{ipag}}
 \and A.~Eckart\inst{\ref{cologne},\ref{bonn}}
 \and F.~Eisenhauer\inst{\ref{mpe}}
 \and N.~M.~F\"orster Schreiber\inst{\ref{mpe}}
 \and P.~Garcia\inst{\ref{centra},\ref{porto}}
 \and R.~Garcia~Lopez\inst{\ref{dublin},\ref{mpia}}
 \and T.~Gardner\inst{\ref{exeterAstro}}
 \and E.~Gendron\inst{\ref{lesia}}
 \and R.~Genzel\inst{\ref{mpe},\ref{ucb}}
 \and S.~Gillessen\inst{\ref{mpe}}
 \and J.~H.~Girard\inst{\ref{stsci}}
 \and S.~Grant\inst{\ref{mpe}}
 \and X.~Haubois\inst{\ref{esoc}}
 \and G.~Hei\ss el\inst{\ref{actesa},\ref{lesia}}
 \and T.~Henning\inst{\ref{mpia}}
 \and S.~Hinkley\inst{\ref{exeter}}
 \and S.~Hippler\inst{\ref{mpia}}
 \and M.~Houll\'e\inst{\ref{cotedazur}}
 \and Z.~Hubert\inst{\ref{ipag}}
 \and L.~Jocou\inst{\ref{ipag}}
 \and J.~Kammerer\inst{\ref{esog}}
 \and M.~Keppler\inst{\ref{mpia}}
 \and P.~Kervella\inst{\ref{lesia}}
 \and L.~Kreidberg\inst{\ref{mpia}}
 \and N.~T.~Kurtovic\inst{\ref{mpe}}
 \and A.-M.~Lagrange\inst{\ref{ipag},\ref{lesia}}
 \and V.~Lapeyr\`ere\inst{\ref{lesia}}
 \and J.-B.~Le~Bouquin\inst{\ref{ipag}}
 \and P.~L\'ena\inst{\ref{lesia}}
 \and D.~Lutz\inst{\ref{mpe}}
 \and A.-L.~Maire\inst{\ref{ipag}}
 \and F.~Mang\inst{\ref{mpe}}
 \and G.-D.~Marleau\inst{\ref{duisburg},\ref{tuebingen},\ref{bern},\ref{mpia}}
 \and A.~M\'erand\inst{\ref{esog}}
 \and J.~D.~Monnier\inst{\ref{umich}}
 \and C.~Mordasini\inst{\ref{bern}}
 \and D.~Mouillet\inst{\ref{ipag}}
 \and E.~Nasedkin\inst{\ref{mpia}}
 \and T.~Ott\inst{\ref{mpe}}
 \and G.~P.~P.~L.~Otten\inst{\ref{sinica}}
 \and C.~Paladini\inst{\ref{esoc}}
 \and T.~Paumard\inst{\ref{lesia}}
 \and K.~Perraut\inst{\ref{ipag}}
 \and G.~Perrin\inst{\ref{lesia}}
 \and O.~Pfuhl\inst{\ref{esog}}
 \and N.~Pourr\'e\inst{\ref{ipag}}
 \and L.~Pueyo\inst{\ref{stsci}}
 \and D.~C.~Ribeiro\inst{\ref{mpe}}
 \and E.~Rickman\inst{\ref{esa}}
 \and Z.~Rustamkulov\inst{\ref{jhueps}}
 \and J.~Shangguan\inst{\ref{mpe}}
 \and T.~Shimizu \inst{\ref{mpe}}
 \and D.~Sing\inst{\ref{jhupa},\ref{jhueps}}
 \and J.~Stadler\inst{\ref{mpa},\ref{origins}}
 \and T.~Stolker\inst{\ref{leiden}}
 \and O.~Straub\inst{\ref{origins}}
 \and C.~Straubmeier\inst{\ref{cologne}}
 \and E.~Sturm\inst{\ref{mpe}}
 \and M. Subroweit\inst{\ref{cologne}} 
 \and L.~J.~Tacconi\inst{\ref{mpe}}
 \and E.F.~van~Dishoeck\inst{\ref{leiden},\ref{mpe}}
 \and A.~Vigan\inst{\ref{lam}}
 \and F.~Vincent\inst{\ref{lesia}}
 \and S.~D.~von~Fellenberg\inst{\ref{bonn}}
 \and J.~J.~Wang\inst{\ref{northwestern}}
 \and F.~Widmann\inst{\ref{mpe}}
 \and T.~O.~Winterhalder\inst{\ref{esog}}
 \and J.~Woillez\inst{\ref{esog}}
 \and \c{S}.~Yaz\i{}c\i{}\inst{\ref{mpe}}
 \and A.~Young\inst{\ref{stockholm}}
 \and  the GRAVITY Collaboration}
\institute{ 
  Institute of Astronomy, University of Cambridge, Madingley Road, Cambridge CB3 0HA, United Kingdom
\label{cam}      \and
   Kavli Institute for Cosmology, University of Cambridge, Madingley Road, Cambridge CB3 0HA, United Kingdom
\label{kavli} \and  
   LESIA, Observatoire de Paris, PSL, CNRS, Sorbonne Universit\'e, Universit\'e de Paris, 5 place Janssen, 92195 Meudon, France
\label{lesia}      \and
   European Southern Observatory, Karl-Schwarzschild-Stra\ss e 2, 85748 Garching, Germany
\label{esog}      \and
   Universidade de Lisboa - Faculdade de Ci\^encias, Campo Grande, 1749-016 Lisboa, Portugal
\label{lisboa}      \and
   CENTRA - Centro de Astrof\' isica e Gravita\c c\~ao, IST, Universidade de Lisboa, 1049-001 Lisboa, Portugal
\label{centra}      \and
   Max Planck Institute for Astronomy, K\"onigstuhl 17, 69117 Heidelberg, Germany
\label{mpia}      \and
   Department of Physics \& Astronomy, Johns Hopkins University, 3400 N. Charles Street, Baltimore, MD 21218, USA
\label{jhupa}      \and
   Space Telescope Science Institute, 3700 San Martin Drive, Baltimore, MD 21218, USA
\label{stsci}      \and
   Univ. Grenoble Alpes, CNRS, IPAG, 38000 Grenoble, France
\label{ipag}      \and
   Center for Interdisciplinary Exploration and Research in Astrophysics (CIERA) and Department of Physics and Astronomy, Northwestern University, Evanston, IL 60208, USA
\label{northwestern}      \and
   Max Planck Institute for extraterrestrial Physics, Giessenbachstra\ss e~1, 85748 Garching, Germany
\label{mpe}      \and
   Aix Marseille Univ, CNRS, CNES, LAM, Marseille, France
\label{lam}      \and
   Université Côte d’Azur, Observatoire de la Côte d’Azur, CNRS, Laboratoire Lagrange, France
\label{cotedazur}      \and
  STAR Institute, Universit\'e de Li\`ege, All\'ee du Six Ao\^ut 19c, 4000 Li\`ege, Belgium
\label{liege}      \and
   Leiden Observatory, Leiden University, P.O. Box 9513, 2300 RA Leiden, The Netherlands
\label{leiden}      \and
   Department of Astrophysical \& Planetary Sciences, JILA, Duane Physics Bldg., 2000 Colorado Ave, University of Colorado, Boulder, CO 80309, USA
\label{boulder}      \and
   1.\ Institute of Physics, University of Cologne, Z\"ulpicher Stra\ss e 77, 50937 Cologne, Germany
\label{cologne}      \and
   Max Planck Institute for Radio Astronomy, Auf dem H\"ugel 69, 53121 Bonn, Germany
\label{bonn}      \and
   Universidade do Porto, Faculdade de Engenharia, Rua Dr.~RobertoRua Dr.~Roberto Frias, 4200-465 Porto, Portugal
\label{porto}      \and
   School of Physics, University College Dublin, Belfield, Dublin 4, Ireland
\label{dublin}      \and
   Astrophysics Group, Department of Physics \& Astronomy, University of Exeter, Stocker Road, Exeter, EX4 4QL, UK
\label{exeterAstro}\and
   Departments of Physics and Astronomy, Le Conte Hall, University of California, Berkeley, CA 94720, USA
\label{ucb}      \and
   European Southern Observatory, Casilla 19001, Santiago 19, Chile
\label{esoc}      \and
   Advanced Concepts Team, European Space Agency, TEC-SF, ESTEC, Keplerlaan 1, NL-2201, AZ Noordwijk, The Netherlands
\label{actesa}      \and
   University of Exeter, Physics Building, Stocker Road, Exeter EX4 4QL, United Kingdom
\label{exeter}\and
   Fakult\"at f\"ur Physik, Universit\"at Duisburg-Essen, Lotharstraße 1, 47057 Duisburg, Germany
\label{duisburg}      \and
   Instit\"ut f\"ur Astronomie und Astrophysik, Universit\"at T\"ubingen, Auf der Morgenstelle 10, 72076 T\"ubingen, Germany
\label{tuebingen}      \and
   Center for Space and Habitability, Universit\"at Bern, Gesellschaftsstrasse 6, 3012 Bern, Switzerland
\label{bern}      \and
   Astronomy Department, University of Michigan, Ann Arbor, MI 48109 USA
\label{umich}      \and
   Academia Sinica, Institute of Astronomy and Astrophysics, 11F Astronomy-Mathematics Building, NTU/AS campus, No. 1, Section 4, Roosevelt Rd., Taipei 10617, Taiwan
\label{sinica}      \and
   European Space Agency, ESA Office, Space Telescope Science Institute, 3700 San Martin Drive, Baltimore, MD 21218, USA
\label{esa}      \and
   Department of Earth \& Planetary Sciences, Johns Hopkins University, Baltimore, MD, USA
\label{jhueps}      \and
   Max Planck Institute for Astrophysics, Karl-Schwarzschild-Str. 1, 85741 Garching, Germany
\label{mpa}      \and
   Excellence Cluster ORIGINS, Boltzmannstraße 2, D-85748 Garching bei München, Germany
\label{origins}      \and
   Department of Astronomy, Stockholm University, AlbaNova University Center, 10691 Stockholm, Sweden
\label{stockholm}    
}

\date{\today}

\abstract
    {Dual-field interferometric observations with VLTI/GRAVITY sometimes require the use of a ``binary calibrator'', a binary star whose individual components remain unresolved by the interferometer, with a separation between 400 and 2000 mas for observations with the Units Telescopes (UTs), or 1200 to 3000 mas for the Auxiliary Telescopes (ATs). The separation vector also needs to be predictable to within $10~\mathrm{mas}$ for proper pointing of the instrument.}
   {Up until now, no list of properly vetted calibrators was available for dual-field observations with VLTI/GRAVITY on the UTs. Our objective is to compile such a list, and make it available to the community.}
   {We identify a list of candidates from the Washington Double Star (WDS) catalogue, all with appropriate separations and brightness, scattered over the Southern sky. We observe them as part of a dedicated calibration programme, and determine whether these objects are true binaries (excluding higher multiplicities resolved interferometrically but unseen by imaging), and extract measurements of the separation vectors. We combine these new measurements with those available in the WDS to determine updated orbital parameters for all our vetted calibrators.}
   {We compile a list of 13 vetted binary calibrators for observations with VLTI/GRAVITY on the UTs, and provide orbital estimates and astrometric predictions for each of them. We show that our list guarantees that there are always at least two binary calibrators at $\mathrm{airmass} < 2$ in the sky over the Paranal observatory, at any point in time.}
   {Any Principal Investigator wishing to use the dual-field mode of VLTI/GRAVITY with the UTs can now refer to this list to select an appropriate calibrator. We encourage the use of "whereistheplanet" to predict the astrometry of these calibrators, which seamlessly integrates with "p2Gravity" for VLTI/GRAVITY dual-field observing material preparation.}


   \maketitle
%
    
\section{Introduction}

\begin{figure*}
  \begin{center}
    \includegraphics[width = \linewidth, trim = 0cm 1cm 0cm 1cm]{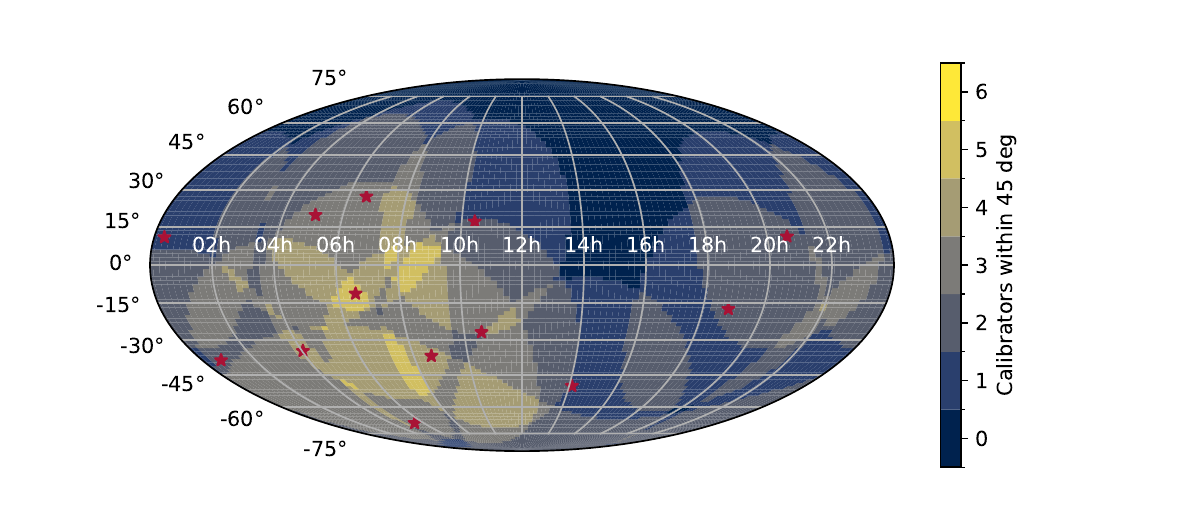}
    \caption{Map representing the on-sky coverage provided by the 13 vetted binary calibrators presented in this work. The shaded regions show the number of available calibrators within $45\,\mathrm{deg}$.}
    \label{fig:map}
  \end{center}
\end{figure*}

Long-baseline optical interferometry is a powerful tool for high-angular resolution astronomy, and the GRAVITY instrument \citep{GRAVITYCollaboration2017} on the Very Large Telescope Interferometer (VLTI) has proven to be extremely useful for the study of many objects, ranging from the centre of our Galaxy \citep{GRAVITYCollaboration2022} and its supermassive black hole \citep{GRAVITYCollaboration2020c, GRAVITYCollaboration2020b, GRAVITYCollaboration2020, GRAVITYCollaboration2018a, GRAVITYCollaboration2018b, GRAVITYCollaboration2019b}, to Active Galactic Nuclei \citep{GRAVITYCollaboration2020} or exoplanets \citep{GRAVITYCollaboration2019a, GRAVITYCollaboration2020d, Nowak2020}.

All these recent achievements have benefited from the unique capability of GRAVITY to obtain high-precision astrometric measurement beyond the diffraction limit of a single telescope of the array, i.e. in the so-called ``dual-field'' mode. In this mode, the two channels of the instrument observe simultaneously two targets: the science target, and a close-by fringe-tracking target. An internal laser metrology system is used to track and correct the non-common path errors between the two channels, so that the relative astrometry between the two targets can be accurately measured. This mode has, among other feats, made GRAVITY successful in high-contrast imaging of exoplanets, and for astrometric monitoring of stars orbiting Sgr A$^*$ in the centre of the Milky Way.

However, in order to derive accurate astrometric measurements from dual-field interferometric observations, a calibration of the metrology reference is needed, which in some situations requires the observation of a dedicated ``binary calibrator''. In the absence of a catalogue of vetted binary calibrators, options were initially very limited. This was a major source of concern during the early days of the ExoGRAVITY Large Programme (1104.C-0651, PI Lacour), and it triggered a series of efforts to assemble such a catalogue, in particular through a dedicated calibration programme (106.21S5.001, PI Nowak).

This paper reports on those efforts, and presents the first catalogue of vetted binary calibrators covering the Southern Sky (Figure~\ref{fig:map}), for VLTI/GRAVITY dual-field observations on the Unit Telescopes (UTs). In Section~\ref{sec:dual-field}, we give some general background on dual-field astrometry with GRAVITY and explain why these binary calibrators are needed in this peculiar observing mode. Section~\ref{sec:assembling} details the procedure we followed to assemble and observe our initial list of targets, and Section~\ref{sec:data-reduction} explains how the data were reduced, and how the calibrators were vetted. In Section~\ref{sec:results} we present the final catalogue, which contain a total of 13 entries. We also report on the numerous astrometric measurements obtained on these targets, and provide orbital parameter estimates and astrometric predictions which can be used for pointing the instrument during observations. Section~\ref{sec:conclusion} gives our final conclusions.

\section{GRAVITY dual-field astrometry and why we need binary calibrators}
\label{sec:dual-field}

\subsection{Optical path, fringe-tracking and metrology in the GRAVITY instrument}

The GRAVITY instrument is built around two different interferometric combiners. The "science" (SC) combiner records the coherent flux at low to medium spectral resolution ($R=50$ to 4000), usually with relatively long integration time (up to 300~s for faint targets). The Fringe-Tracker (FT) combiner enables such long integrations, by actively compensating for the atmospheric turbulence. This second combiner records the coherent flux at much lower resolution and with short integration time. It actively measures and records the phase of the coherent flux, and control a set of actuators to keep the interferometric signal stable \citep[Nowak et al. in prep]{Lacour2019}.

The use of two beam combiners is one of the strengths of GRAVITY, but it also makes it difficult to keep track of the different Optical Paths Length (OPL) in the instrument. If we assume that both the FT and the SC are pointed towards two different close-by target stars $S_\mathrm{FT}$ and $S_\mathrm{SC}$, then for a baseline $(j,k)$ linking telescope $T_j$ to $T_k$, the recorded coherent fluxes in both channels as a function of time and wavelength are given by
\begin{align}
\Gamma^{\mathrm{FT}}_{j,k}(t, \lambda) &=  T(t, \lambda)I_{S_\mathrm{FT}}(\lambda)\exp\left[-i\frac{2\pi}{\lambda}\mathrm{OPD}^\mathrm{FT}_{j,k}(t)\right] \label{eq:basic_FT},\\
\Gamma^{\mathrm{SC}}_{j,k}(t, \lambda) &=  T(t, \lambda)I_{S_\mathrm{SC}}\lambda)\exp\left[-i\frac{2\pi}{\lambda}\mathrm{OPD}^\mathrm{SC}_{j,k}(t)\right].
\label{eq:basic_SC}
\end{align}
\noindent{}In these equations, $I(\lambda)$ represents the spectrum of the target, $T$ the transmission (of the atmosphere and instrument), and $\mathrm{OPD}^\mathrm{SC/FT}_{j,k}$ refers to the effective Optical Path Difference on baseline $(j,k)$ for the given channel. 

\begin{figure*}
    \begin{center}
        \includegraphics[width=0.9\linewidth]{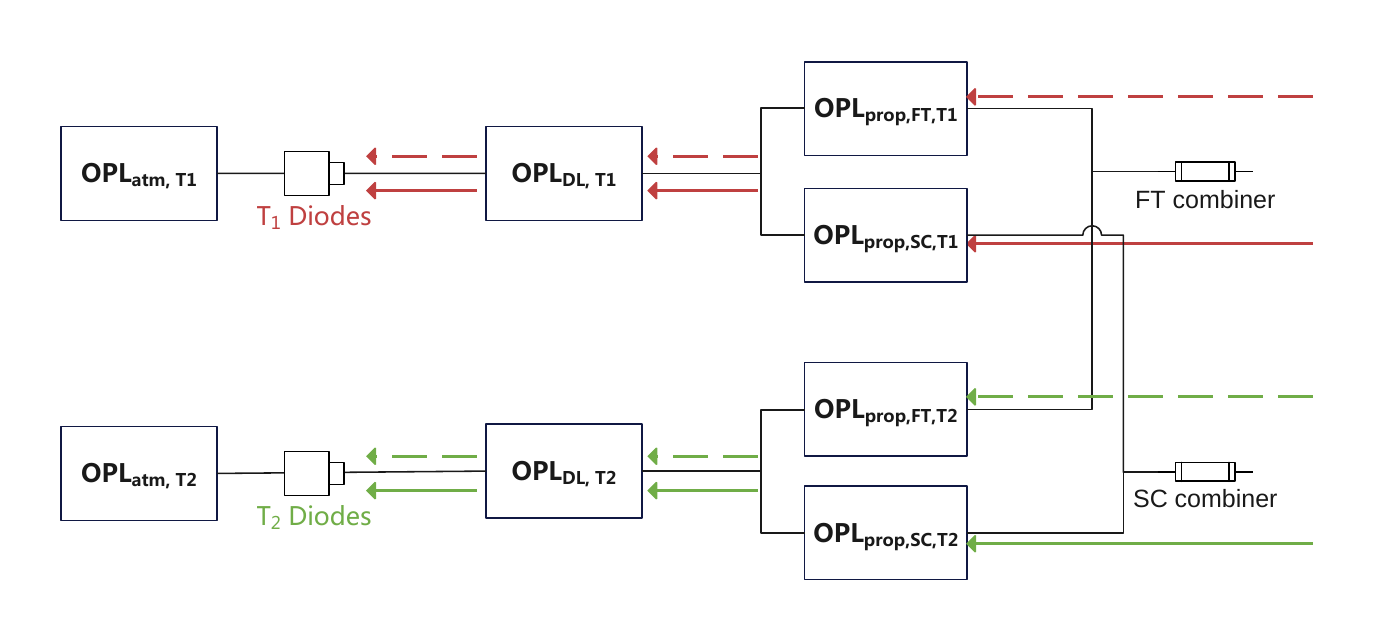}
        \caption{Block diagram showing the different elements contributing to the total OPL within the two channels of the GRAVITY when used in dual-field mode.}
    \label{fig:opd}
    \end{center}
\end{figure*}

The OPD terms in Equations~\ref{eq:basic_FT} and \ref{eq:basic_SC} include contributions from very different elements, which are summarized on the block diagram presented in Figure~\ref{fig:opd}. From the target to the FT or SC channel via telescope $T_k$, the total optical path is the sum of:
\begin{enumerate}
    \item The "in-vacuum" OPL from the FT or SC target to telescope $T_k$, $\mathrm{OPL}^{S_\mathrm{SC/FT}}_k(t)$. This OP depends on time through the rotation of the Earth.
    \item An extra time-dependent contribution from the atmosphere above telescope $T_k$, $\mathrm{OPL}_{\mathrm{atm}, k}(t)$.
    \item A contribution representing the path from telescope $T_k$ through the VLTI Delay Lines (DL) and up to the GRAVITY injection unit (where the beam is split between the two channels). This can be represented as a single telescope dependent quantity $\mathrm{OPL}_{\mathrm{DL},k}(t)$, and is the quantity actively controlled by the Fringe-Tracker. 
    \item A channel dependent contribution which represents the path from the injection unit for the beam coming from $T_k$ to the detector, through the given channel (SC/FT): $\mathrm{OPL}^\mathrm{FT/SC}_{k}(t)$ 
\end{enumerate}

The total OPL from the target to the detector of channel FT or SC is therefore written
\begin{equation}
    \begin{split}
    \mathrm{OPL}^\mathrm{SC/FT}_{\mathrm{tot}, k}(t) =  \mathrm{OPL}^{S_\mathrm{SC/FT}}_k(t) &+ \mathrm{OPL}_{\mathrm{atm}, k}(t) \\ &+ \mathrm{OPL}_{\mathrm{DL},k}(t) + \mathrm{OPL}^\mathrm{FT/SC}_{k}(t).
    \end{split}
\end{equation}
\noindent{}Using this expression, the OPDs on baseline $(j,k)$ seen by the two beam combiners are given by
\begin{equation}
    \begin{split}
    \mathrm{OPD}^\mathrm{SC/FT}_{\mathrm{tot}, j, k}(t) = \mathrm{OPD}^{S_\mathrm{SC/FT}}_{j, k}(t) &+ \mathrm{OPD}_{\mathrm{atm}, {j, k}}(t) \\ &+ \mathrm{OPD}_{\mathrm{DL}, j, k}(t) + \mathrm{OPD}^\mathrm{FT/SC}_{j, k}(t),
    \end{split}
    \label{eq:OPD}
\end{equation}
\noindent{}where any $\mathrm{OPD}_{j, k}$ term is to be understood as the difference $\mathrm{OPL}_k - \mathrm{OPL}_j$. 

From an astrophysical perspective, the quantity of interest in Equation~\ref{eq:OPD} is the "in-vacuum" term $\mathrm{OPD}^{S_\mathrm{SC}}_{j, k}$, which is directly related to the astrometric position of the science target $S_\mathrm{SC}$. To extract this astrometry, one of the main steps performed during the data reduction is to "reference the science channel to the FT channel". What this really means is that the phase of the FT coherent flux is extracted, interpolated on to the SC wavelength grid, and subtracted from the SC measurements, according to
\begin{equation}
\begin{split}
\tilde{\Gamma}^{\mathrm{SC}}_{j,k}(t, \lambda) &= \Gamma^{\mathrm{SC}}_{j,k}(t, \lambda)\times{}\exp\left\{i\,\arg\left[\Gamma^{\mathrm{FT}}_{j,k}(t, \lambda)\right]\right\} \\
&=T(t, \lambda)I_{S_\mathrm{SC}}(\lambda)\\
&\qquad\times{}\exp\left(-i\frac{2\pi}{\lambda}\left[\mathrm{OPD}^{S_\mathrm{SC}}_{j,k}(t) - \mathrm{OPD}^{S_\mathrm{FT}}_{j,k}(t)\right]\right) \\
&\qquad\times{}\exp\left(-i\frac{2\pi}{\lambda}\left[\mathrm{OPD}^\mathrm{SC}_{j,k}(t) - \mathrm{OPD}^\mathrm{FT}_{j,k}(t)\right]\right).
\end{split}
\label{eq:Gamma_tilde}
\end{equation}

The expression of $\tilde{\Gamma}^{\mathrm{SC}}_{j,k}$ shows two remaining differential OPD terms. The first one is the difference between the "in-vacuum" OPD on baseline $(j,k)$ for the SC target and the FT target. This is a purely astrophysical quantity, which is directly related to the differential astrometry $\Delta{}\mathrm{RA}$ and $\Delta{}\mathrm{DEC}$ between the SC and FT targets, and to the baseline $(u, v)$ coordinates:
\begin{equation}
    \left[\mathrm{OPD}^{S_\mathrm{SC}}_{j,k}(t) - \mathrm{OPD}^{S_\mathrm{FT}}_{j,k}(t)\right] = \left[\Delta\mathrm{RA}\times{}u(t) + \Delta\mathrm{DEC}\times{}v(t)\right].
\end{equation}
\noindent{}On the opposite, the second differential term is a purely instrumental quantity. It is better understood by reordering the terms in the difference using
\begin{small}
\begin{equation}
\begin{split}
    \left[\mathrm{OPD}^{\mathrm{SC}}_{j,k}- \mathrm{OPD}^{\mathrm{FT}}_{j,k}\right] &= (\mathrm{OPL}^{\mathrm{SC}}_{k} - \mathrm{OPL}^{\mathrm{SC}}_{j}) - (\mathrm{OPL}^{\mathrm{FT}}_{k} - \mathrm{OPL}^{\mathrm{FT}}_{j}) \\
    &= (\mathrm{OPL}^{\mathrm{SC}}_{k} - \mathrm{OPL}^{\mathrm{FT}}_{k}) - (\mathrm{OPL}^{\mathrm{FT}}_{j} - \mathrm{OPL}^{\mathrm{SC}}_{j}) \\
    &= \Delta{}\mathrm{OPL}^{\mathrm{SC-FT}}_{k} - \Delta{}\mathrm{OPL}^{\mathrm{SC-FT}}_{j}
\end{split}
,\end{equation}\end{small}\noindent{}which demonstrates that this instrumental part stems from a non common-path error between the FT and SC channels. 

In GRAVITY, this non common-path error is measured by a laser-metrology system \citep{Gillessen2012}. For this, a laser beam is sent backwards in the optical train, from the beam-combiner to the telescopes. In the pupil-plane of each telescope, the laser light from both the SC and FT channels interfere to create fringes, which are recorded by a set of diodes on the telescope spider. From this, each of the four $\Delta{}\mathrm{OPL}^{\mathrm{SC-FT}}$ terms (one per telescope) can be extracted. This allows for the correction of this term, up to an unknown wavelength-dependent "metrology zero-point" $\phi^\mathrm{ZP}(\lambda)$ which stems from the uncontrolled injection phase of the laser beam. During the data-reduction, for any baseline $(j, k)$, both $\Delta{}\mathrm{OPL}^{\mathrm{SC-FT}}_{k}$ and $\Delta{}\mathrm{OPL}^{\mathrm{SC-FT}}_{j}$ can be computed, and removed from the SC coherent flux. This leaves the "metrology-corrected" coherent flux, expressed as
\begin{equation}
  \begin{split}
    \Gamma^\mathrm{metcorr}{j, k}(t, \lambda) = T(t, \lambda)&I_{S_\mathrm{SC}}(\lambda)\\
    & \times{} e^{-i\frac{2\pi}{\lambda}\left[\Delta\mathrm{RA}\times{}u(t)+\Delta\mathrm{DEC}\times{}v(t)\right]-i\phi^\mathrm{ZP}_{j,k}(\lambda)}.
    \label{eq:Gamma_metcorr}
  \end{split}
\end{equation}

Equation~\ref{eq:Gamma_metcorr} shows how the metrology zero-point can interfere with the measurement of the astronomy of the target. As such, one of the main concerns of the dual-field observing strategy and the astrometric extraction is to find a way to determine and correct this unknown term $\phi_\mathrm{ZP}(\lambda)$.

\subsection{First observing strategy: on-axis}
\label{sec:on-axis}

A first option to determine the metrology zero-point (also known as the "phase reference") is to point both the SC and FT channels at the same target. In this case, the relative astrometry between the channels is zero, and the terms $\Delta\mathrm{RA}$ and $\Delta\mathrm{DEC}$ vanish from Equation~\ref{eq:Gamma_metcorr}. The coherent flux on the SC channel after correction from the FT phase and the metrology is almost a direct measurement of $\phi_\mathrm{ZP}$:
\begin{equation}
    \Gamma^\mathrm{metcorr}_{j, k}(t, \lambda) = T(t, \lambda)I_{S_\mathrm{SC}}(\lambda)e^{-i\phi^\mathrm{ZP}_{j,k}(\lambda)}.
\end{equation}

\begin{figure*}
    \begin{center}
        \includegraphics[width=0.9\linewidth]{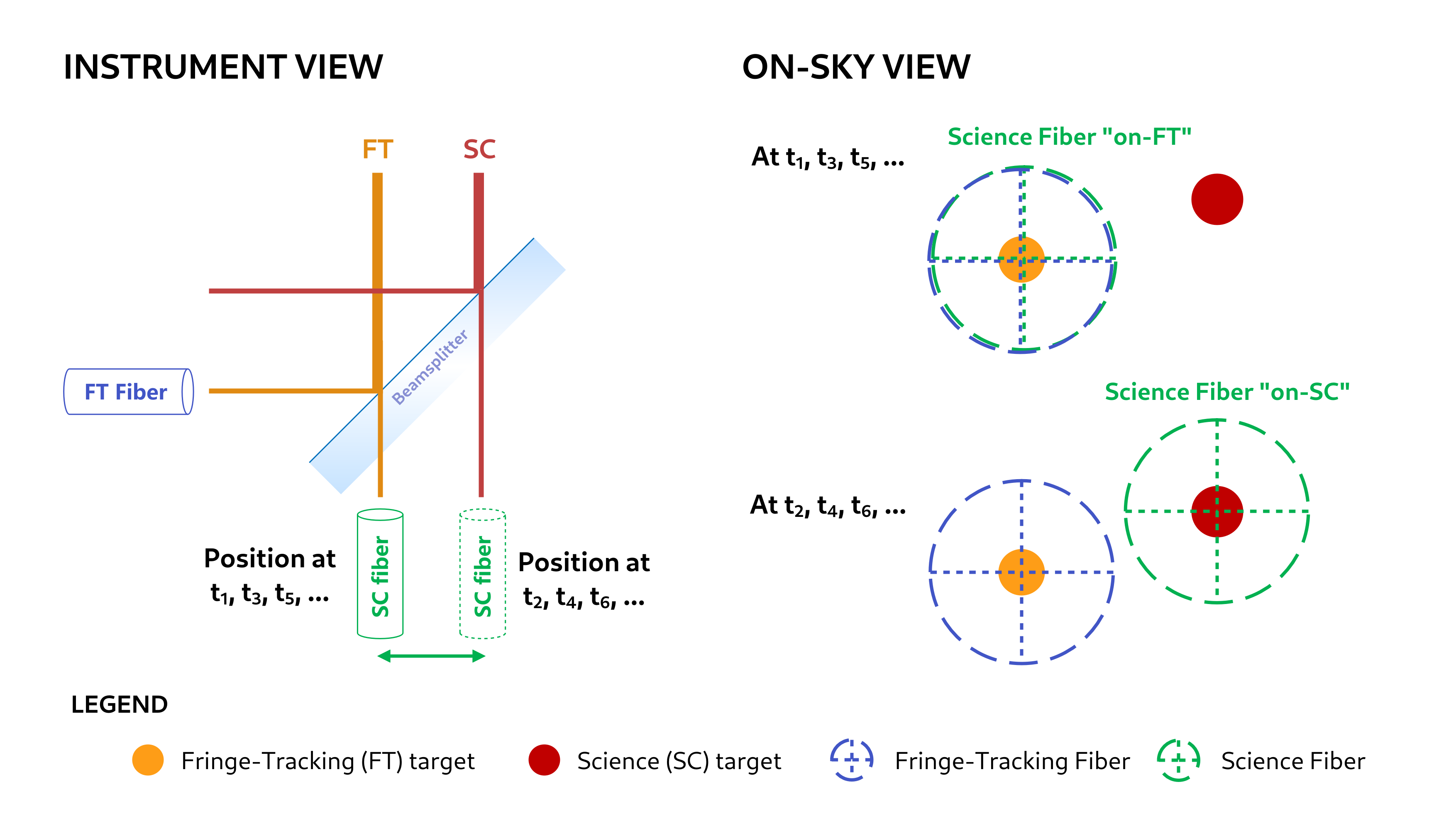}
        \caption{Illustration of the ``dual-field on-axis'' observing strategy. This strategy is used to measure the astrometry of a scientific target relative to a fringe-tracking target. The left panel illustrates the position of the two fibres in the instrument, with the use of a beam-splitter to separate the field. The right panel illustrates the resulting position of the two fibres on-sky.}
    \label{fig:onaxis}
    \end{center}
\end{figure*}

This forms the basis of the on-axis observing strategy, which consists in interleaving observations with the SC channel pointed on the science target with observations with the SC channel pointed at the FT target (see Figure~\ref{fig:onaxis}). For an on-science observation obtained at time $t$, the phase zero-point is obtained from the average of the on-FT observation obtained just before (at $t_1<t$), and the on-FT observation obtained just after (at $t_2>t$) according to

\begin{equation}
    \phi^\mathrm{ZP}_{j, k}(\lambda) = -\frac{\arg\left\{\Gamma^\mathrm{metcorr}_{j,k}(t_1, \lambda)\right\} + \arg\left\{\Gamma^\mathrm{metcorr}_{j,k}(t_2, \lambda)\right\}}{2},
    \label{eq:zp_onaxis}
\end{equation}
\noindent{}and the "phase-referenced" coherent flux is calculated using
\begin{equation}
    \Gamma^\mathrm{ref}_{j,k}(t, \lambda) = \Gamma^\mathrm{metcorr}_{j,k}(t, \lambda)e^{i\phi^\mathrm{ZP}_{j,k}(\lambda)}.
\end{equation}

The amplitude of this phase-referenced coherent flux is determined only by the spectrum of the science target and by the combined atmosphere-instrument transmission, whereas its phase is determined only by the relative astrometry with respect to the FT target. In other words, we have:
\begin{equation}
    \Gamma^\mathrm{ref}_{j,k}(t, \lambda) = T(t, \lambda)S_{S_\mathrm{SC}}(\lambda)e^{-i\frac{2\pi}{\lambda}\left[\Delta{}\mathrm{RA}\times{}u(t)+\Delta\mathrm{DEC}\times{}v(t)\right]}.
    \label{eq:Gamma_ref}
\end{equation}

From a technical perspective, pointing the FT and SC channels simultaneously towards the same target to extract the metrology zero-point requires the use of a beam-splitter to separate the beam in amplitude. The phase-reference remains valid as long as the instrumental setup does not change, which means that the beam-splitter must also be used for the "on-science" observations. In GRAVITY, the beam-splitter can only accommodate a maximum separation between the FT and SC fibres of $\sim{}500\,\mathrm{mas}$ and thus, beyond this, a different strategy is required. The beam-splitter also reduces the flux by a factor 2, which might not be ideal for fainter targets.

\subsection{Second observing strategy: off-axis swap}
\label{sec:off-axis}

When the separation between the fringe-tracking and science targets is too large to use the beam-splitter, the only remaining option in GRAVITY is to use a roof-mirror to split the field between the SC and FT channels. However, in this case, the two channels cannot physically be pointed simultaneously at the same target to obtain the metrology zero-point. Another strategy must therefore be used to determine the metrology zero-point in Equation~\ref{eq:Gamma_metcorr}. 

If the science target itself is suitable for fringe-tracking, a possibility is to perform a "swap" observation. In this case, the two fibres are "swapped". This reverses the differential astrometry in Equation~\ref{eq:Gamma_metcorr}, and results in the measurement of two coherent fluxes (at $t_{-}$ before the swap, and $t_{+}$ after the swap) given by:
\begin{align}
    \Gamma^\mathrm{metcorr}_{j, k}(t_-, \lambda) &= T(t_-, \lambda)S(\lambda)e^{-i\frac{2\pi}{\lambda}\left[u(t_-)\Delta\mathrm{RA}+v(t_-)\Delta\mathrm{DEC}\right]-i\phi^\mathrm{ZP}_{j,k}(\lambda)}, \\
    \Gamma^\mathrm{metcorr}_{j, k}(t_+, \lambda) &= T(t_+, \lambda)S(\lambda)e^{+i\frac{2\pi}{\lambda}\left[u(t_+)\Delta\mathrm{RA}v(t_+)+\Delta\mathrm{DEC}\right]-i\phi^\mathrm{ZP}_{j,k}(\lambda)},
\end{align}
\noindent{}and from which the metrology zero-point can be extracted using
\begin{equation}
    \phi^\mathrm{ZP}_{j, k}(\lambda) = -\frac{\arg\left\{\Gamma^\mathrm{metcorr}_{j,k}(t_-, \lambda)\right\} + \arg\left\{\Gamma^\mathrm{metcorr}_{j,k}(t_+, \lambda)\right\}}{2}.
    \label{eq:zp_offaxis}
\end{equation}

\begin{figure*}
    \begin{center}
        \includegraphics[width=0.9\linewidth]{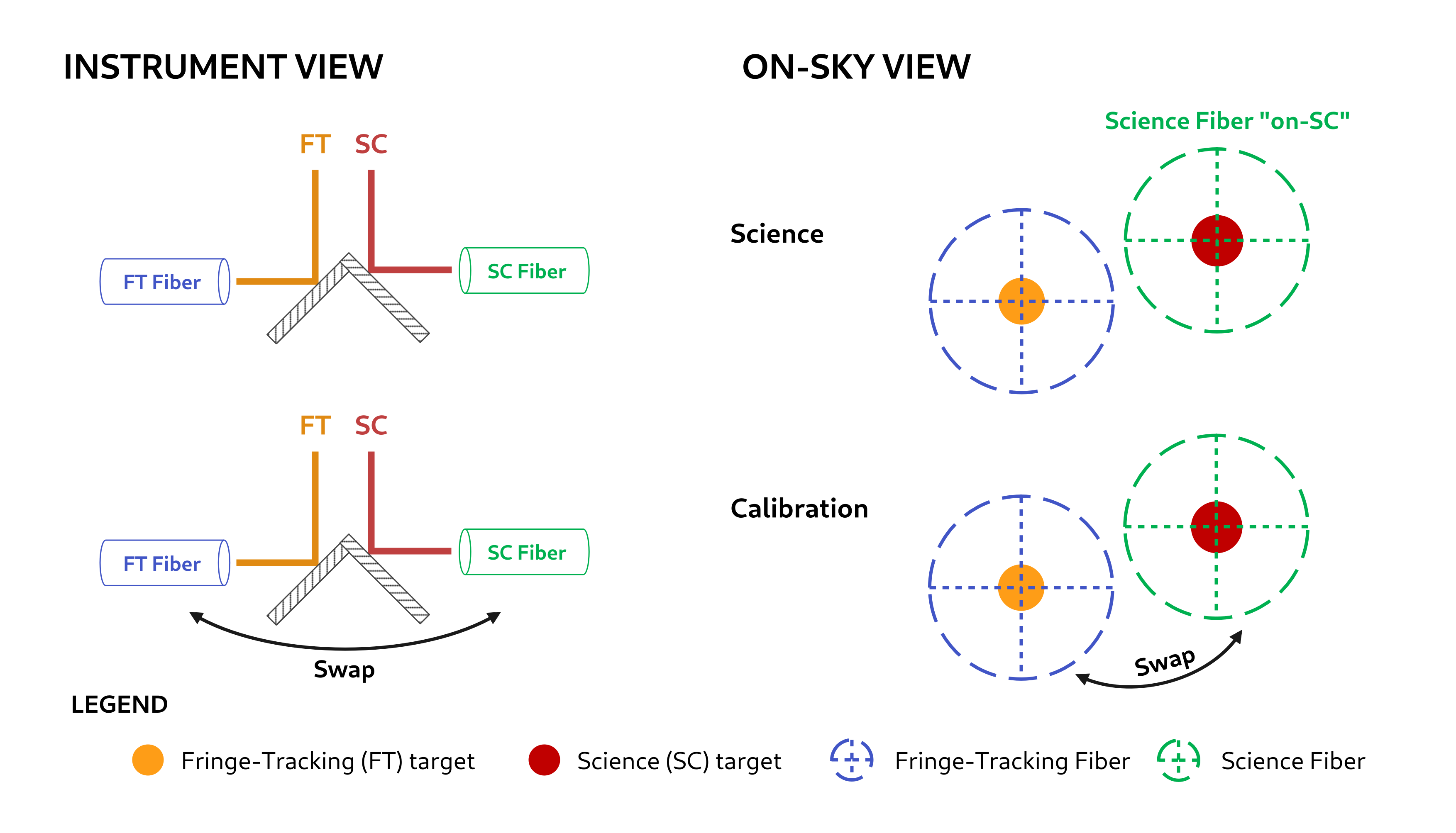}
        \caption{Similar to Figure~\ref{fig:onaxis}, but illustrating the ``dual-field off-axis'' observing strategy. In this case, a roof-mirror is used to separate the field, and two sequences are required. In the ``science'' sequence, the FT fibre is positioned on an appropriately selected FT target, whereas the SC fibre is centred on the science target. In the ``calibration'' sequence, a binary star is observed multiple times, while swapping the position of the two fibres. If the science target is appropriate for fringe-tracking, the calibration can be performed on the science target itself. If not, the calibration sequence must be performed on a separate binary calibrator must be observed.}
        \label{fig:offaxis}
        \end{center}
\end{figure*}

The coherent flux at both $t_+$ and $t_-$ can then be corrected for this zero-point, to yield a phase-referenced coherent-flux similar to Equation~\ref{eq:Gamma_ref}. Note that in this scenario, represented in Figure~\ref{fig:offaxis}, half of the observations are measuring the astrometry of the science target relative to the FT target, whereas the other half is measuring the astrometry of the FT target relative to the main scientific target. 

\subsection{The need for binary calibrators}

For the off-axis swap strategy to work, the science target itself needs to be suitable for fringe-tracking. When it is not (too faint, for example), the swap strategy cannot be performed on the FT-SC pair. In this case, the remaining option is to perform a dedicated dual-field calibration observations. This involves moving the telescopes towards a "binary calibrator", and performing the swap observation on this binary target. The metrology zero-point is extracted from the calibration observation, and subtracted from the science observations. 

To perform this separate calibration observation, a suitable binary calibrator must be available. A proper binary calibrator for an observation with GRAVITY on the UTs is typically defined as a visual binary with a separation ranging from $\sim{}400~\mathrm{mas}$ (to easily separate the two components with the individual telescopes) to $2000~\mathrm{mas}$ (to stay within the available field-of-view of GRAVITY), for which the two components must be of a least of magnitude 8.5 in K-band (for fringe-tracking even in poor conditions) and unresolved by the interferometer.

\section{Assembling the catalogue: target selection and observations}
\label{sec:assembling}

  Three binary calibrators were known at the start of the Exo\-GRAVITY Large Program: WDSJ04021-3429, WDSJ14077-4952, and WDSJ20399+1115. They did not cover the entire Southern Sky, and in an effort to expand on this very limited list, we looked for more candidates in the Washington Double Star (WDS) catalogue.  
  
  We assembled a list of binary calibrator candidates by selecting targets from the WDS catalogue using specific criteria. These criteria included a separation ranging from 400 mas to 2000~mas, and a K band magnitude below 8 to ensure that a good SNR can be reached with minimal observing time. We also looked for targets with available orbit predictions for proper pointing of the single-mode fibres during observations, and selected a sample covering the entire Southern Sky. We did not aim for completeness in our selection.

  From the pool of potential targets, a subset of 21 binaries was chosen to provide broad coverage across the Southern sky. A calibration program was submitted to ESO, and the selected targets were observed with the Auxiliary Telescopes (ATs). The original observation plan spanning P106 to P107 was adjusted due to COVID-19 disruptions, leading to observations being conducted exclusively during the Chilean summer (P106 and P108). Consequently, most targets within RA range of 10h 00min 00s to 20h 00min were regrettably not observed (see Table~\ref{tab:selection}). Thankfully, one of the three original calibrators is located at 14h of RA, which limits the impact of this disruption.

  Observations were conducted using the GRAVITY instrument on the ATs, in its dual-field on-axis mode (see Section~\ref{sec:on-axis} and Figure~\ref{fig:onaxis}). In addition to these observations, dual-field calibration observations from the ExoGRAVITY Large Program and other related programmes were also incorporated, in an effort to improve the orbital parameter estimates. These calibration observations were originally acquired using the GRAVITY instrument on the UTs in the off-axis dual-field mode, with the ``swap'' strategy (see Section~\ref{sec:off-axis} and Figure~\ref{fig:offaxis}). The observing logs describing the data included in this study are given in Appendix~\ref{app:logs}.
  
\begin{table*}
  \begin{center}
    \begin{tabular}{l l l l l l l c}
      \hline
      \hline
      WDS identifier & Alternative name & RA & DEC & \multicolumn{2}{c}{K-band magnitudes} & Separation & Observed$^\dagger$ \\
                     & & [hr:min:sec] & [deg:min:sec] & Primary & Secondary & [mas] &  \\
      \hline
      WDSJ00209+1059 & HD 1663 & 00:20:55 & +10:58:36 & 6.2 & 8.0 & 774 & \cmark \\
      WDSJ00427-3828 & $\lambda$ 01 Scl & 00:42:43 & -38:27:49 & 6.1 & 6.5 & 543 & \cmark \\
      WDSJ02407+2637 & HD 16638 & 02:40:42 & +26:37:20 & 6.2 & 7.4 & 578 & \cmark \\
      WDSJ05055+1948 & HD 32642 & 05:05:32 & +19:48:24.2 & 6.0 & 6.5 & 967 & \cmark \\
      WDSJ05270-6837 & HD 36584 & 05:26:59 & -68:37:21.1 & 5.1 & 5.4 & 1487 & \cmark \\
      WDSJ05417-0254 & HD 37904 & 05:41:40 & -02:53:47.2 & 5.6 & 7.1 & 542 & \cmark \\
      WDSJ06345-1114 & HD 46716 & 06:34:32 & -11:13:47.0 & 6.3 & 7.7 & 693 & \cmark \\
      WDSJ06364+2717 & HD 46780 & 06:36:26 & +27:16:42.2 & 5.3 & 7.2 & 731 & \cmark \\
      WDSJ08394-3636 & HD 73900 & 08:39:22 & -36:36:24.0 & 5.0 & 6.5 & 887 & \cmark \\
      WDSJ10269+1713 & HD 90444 & 10:26:53 & +17:13:10.0 & 6.2 & 6.9 & 802 & \cmark \\
      WDSJ10361-2641 & HD 91881 & 10:36:04 & -26:40:31.7 & 5.1 & 6.1 & 1330 & \xmark \\
      WDSJ10426+0335 & 34 Sex & 10:42:37 & +03:34:59.1 & 5.3 & 6.8 & 619 & \cmark \\
      WDSJ14534+1542 & HD 131473 & 14:53:23 & +15:42:18.5 & 5.1 & 5.7 & 946 & \xmark \\
      WDSJ15557-2645 & HD 142456 & 15:55:41 & -26:44:49.9 & 6.2 & 6.6 & 596 & \xmark \\
      WDSJ16514+0113 & 21 Oph & 16:51:24 & +01:12:57.4 & 5.3 & 6.8 & 767 & \xmark \\
      WDSJ17443-7213 & HD 159964 & 17:44:19 & -72:13:15.3 & 5.2 & 6.4 & 752 & \xmark \\
      WDSJ18236-2610 & HD 168991 & 18:23:33 & -26:10:17.6 & 6.4 & 8.0 & 528 & \xmark \\
      WDSJ18250-0135 & HD 169493 & 18:24:57 & -01:34:45.8 & 5.1 & 5.6 & 906 & \xmark \\
      WDSJ18320+0647 & HD 170987 & 18:32:01 & +06:46:48.3 & 6.3 & 7.4 & 566 & \xmark \\
      WDSJ19346+1808 & HD 184591 & 19:34:37 & +18:07:41.2 & 5.1 & 6.2 & 598 & \xmark \\
      WDSJ22180-6249 & HD 211299 & 22:18:02 & -62:48:40.7 & 5.8 & 6.9 & 682 & \cmark \\
      \hline
    \end{tabular}
    \caption{Initial list of target selected from the WDS catalogue. \\
    $\dagger$ Due to COVID-19 related disruptions at the Paranal observatory, not all targets have been observed.}
    \label{tab:selection}
  \end{center}
\end{table*}

\section{Data-reduction}
\label{sec:data-reduction}

\subsection{Pipeline reduction and general steps}

The first step in our data reduction was to use the ESO GRAVITY pipeline \citep{Lapeyrere2014} to create the intermediate "astroreduced" files, in which the individual detector integrations (DITs) are not averaged. The GRAVITY pipeline can also produce fully reduced "dualscivis" files in which the DITs are averaged. However, as we explain in Section~\ref{sec:astrometry_offaxis}, averaging coherent flux without blurring the signal requires some careful considerations about the astrometry of the targets. The GRAVITY pipeline assumes that the fibre is well-centred on the target, and uses the coordinates of the fibre as the assumption on the astrometry required to calculate the average. In many of our observations, the coordinates of the secondary were poorly known, so this assumption was not always valid. Therefore, in all this work, we decided to keep the "astroreduced" data, and perform any required DITs averaging manually, in our own astrometry extraction routines. 

As a note, the fringe-tracker referencing and metrology correction described in Equation~\ref{eq:Gamma_tilde} and \ref{eq:Gamma_metcorr} are not performed by the pipeline on the astroreduced data product. Therefore, we included these steps in our own routines, using the algorithms described in the GRAVITY Pipeline User manual (VLT-MAN-ESO-19500-XXXX).

\subsection{Vetting of the AT observations}
\label{sec:vetting_ATs}

Our analysis of the binary calibrator candidates started by a check for potential higher-multiplicity systems. A binary calibrator needs to be a proper binary for it to be useful for extracting the metrology zero-point, otherwise, the phase of Equation~\ref{eq:Gamma_metcorr} is affected by an additional term coming from the resolved system, which cannot be easily accounted for in all the subsequent calculations. 

To determine whether the two components of any given target binary are seen as point sources (i.e. are unresolved) by the VLTI, we treated each of them independently, in a similar fashion as if these were "single-field" observations, and performed a $\chi^2$-test. This starts by calculating the interferometric visibility, defined by
\begin{equation}
    V_{j, k}(t, \lambda) = \frac{\Gamma^\mathrm{metcorr}_{j, k}(t, \lambda)}{\sqrt{F_j(t, \lambda)\times{}F_k(t, \lambda)}},
\end{equation}
\noindent{}where $F_i$ refers to the flux on telescope $i$. In theory, for an unresolved component, the amplitude of this visibility should remain constant and equal to 1, and the closure-phase calculated from this visibility should remain equal to 0. In practice, though, the amplitude of the visibility is never equal to 1, and needs to be calibrated with the observation of a known unresolved point-source (an "interferometric calibrator"). As we did not obtain such calibrations during our observations, we focused on the closure-phase to determine whether the individual components of our target binaries are unresolved. 

For a given triangle of 3 telescopes $(j, k, l)$, we calculated the bispectrum using
\begin{equation}
    B_{j, k, l} = \Gamma^\mathrm{metcorr}_{j, k}\times{}\Gamma^\mathrm{metcorr}_{k, l}\times{}\mathrm{conj}\left\{\Gamma^\mathrm{metcorr}_{j, l}\right\},
\end{equation}
\noindent{}and then extracted the closure phase, defined as the argument of this bispectrum:
\begin{equation}
    \Phi_{j, k, l} = \arg(B_{j, k, l}).
\end{equation}
\noindent{}The error bars $\sigma_{j, k, l}$ on the closure phase were propagated from the error bars reported by the pipeline on the coherent fluxes. We then performed a detrending step by fitting and subtracting a 2nd-order polynomial in wavelength for each exposure and triangle. If the component is truly unresolved, the resulting closure phase should be 0, within the error bars. Therefore, we calculate the reduced $\chi^2$ under this null-hypothesis:
\begin{equation}
    \chi_\mathrm{red}^2 = \frac{1}{\mathrm{d.o.f.}}\sum_{\substack{\lambda, t \\ (j, k, l)}}\left[\frac{\Phi_{j, k, l}(t, \lambda)}{\sigma_{j, k, l}(t, \lambda)}\right]^2.
\end{equation} 
\noindent{}In this equation, $\mathrm{d.o.f.}$ stands for the number of degrees of freedom and is given by
\begin{equation}
    \mathrm{d.o.f.} = 4\times{}n_t\times{}(n_\lambda-3),
\end{equation}
\noindent{}since 4 corresponds to the number of independent triangles with 6 baselines, $n_\lambda$ to the number of wavelength channels, $n_t$ to the number of exposures, and "minus 3" accounts for the 3 coefficients of the detrending polynomial.

The values of these reduced $\chi_\mathrm{red}^2$ are reported in Table~\ref{tab:vetted}. Their distribution shows a very clear separation between the unresolved sources, with values close to 1.3 to 1.8, and resolved sources, where the $\chi_\mathrm{red}^2$ spikes to values of $>7$. We adopt a threshold of 2, and reject any candidate binary with one component resulting in a $\chi_\mathrm{red}^2 > 5$ from our list.

\subsection{Extracting the astrometry}

\subsubsection{On-axis observations}

The extraction of astrometric measurements from a sequence of on-axis observations with exposures on the science target interleaved with exposures on the FT target has been described extensively in the frame of exoplanet observations\citep{GRAVITYCollaboration2020, Nowak2020}. In the case of the binary stars presented in this work, there are a few minor differences which need consideration. First, at the level of contrasts and separations considered in this work, the contamination of on-science observations with FT star residuals is not really an issue, meaning that when the SC fibre is centred on one of the two components, the other component does not significantly contribute to the coherent flux. Second, we are only interested in extracting the astrometry, and not the contrast spectrum between the two components. 

With these difference in mind, we follow a similar set of steps as the one described in the above-mentioned papers. We first combined all the on-FT coherent fluxes to extract the metrology zero-point:
\begin{align}
    \phi_{j, }^\mathrm{ZP}(\lambda) &= \arg\left(\left<\Gamma_{j, k}^\mathrm{ref}(t, \lambda)\right>_\mathrm{on-FT}\right).
\end{align}
\noindent{}Since we were not interested in extracting a proper contrast spectrum, and as only one component contributes to the coherent flux in these low-contrast observation, the amplitude reference was simply taken as the modulus of the coherent flux, using
\begin{equation}
    A_{j, k}(t, \lambda) = \left|\Gamma^\mathrm{metcorr}_{j, k}(t, \lambda)\right|.
\end{equation}
\noindent{}From this, we defined a simple binary star model using the metrology zero-point, the amplitude reference, and the binary astrometry $(\Delta\mathrm{RA}, \Delta\mathrm{DEC})$:
\begin{equation}
    \begin{split}
    \Gamma^\mathrm{model}_{j, k}(\Delta\mathrm{RA}, \Delta\mathrm{DEC}, t, \lambda) = & A_{j, k}(t, \lambda)e^{i\phi^\mathrm{ZP}_{j, k}(\lambda)}\\ & \times{}e^{-i\frac{2\pi}{\lambda}\left[u(t)\Delta\mathrm{RA}+v(t)\Delta\mathrm{DEC}\right]},
    \end{split}
\end{equation}
\noindent{}and defined the $\chi^2$ associated to an on-science exposure as
\begin{small}
\begin{equation}
    \chi^2\begin{pmatrix}\Delta\mathrm{RA} \\ \Delta\mathrm{DEC}\end{pmatrix} = \sum_{\substack{(j, k) \\ t, \lambda}}\left(\frac{\Gamma^\mathrm{metcorr}_{j, k}(t, \lambda) - \Gamma^\mathrm{model}_{j, k}(\Delta\mathrm{RA}, \Delta\mathrm{DEC}, t, \lambda)}{\sigma_{j, k}(t, \lambda)}\right)^2,
\end{equation}
\end{small}\noindent{}where the summation runs over all baselines $(j, k)$, all wavelength channels $\lambda$, and all DITs $t$ within the exposure file.

The corresponding best estimate of the astrometry was taken as the location of the minimum $\chi^2$. This resulted in a series of measurements $(\Delta\mathrm{RA}, \Delta\mathrm{DEC})^T_k$, with $1<k<n_\mathrm{exp}$, $n_\mathrm{exp}$ being the number of observations taken with the SC fibre on the secondary component ("on-science" observations, as opposed to "on-FT" where the SC fibre is on the same component as the FT fibre). The final best estimate of the astrometry, with is associated covariance matrix, was taken as the mean and covariance of the series of individual measurement:
\begin{small}
   \begin{align}
    &\begin{pmatrix}\Delta\mathrm{RA} \\ \Delta\mathrm{DEC}\end{pmatrix} = \frac{1}{n_\mathrm{exp}}\sum_k\begin{pmatrix}\Delta\mathrm{RA} \\ \Delta\mathrm{DEC}\end{pmatrix}_k, \\
    &W = \begin{bmatrix}\sigma_{\Delta\mathrm{RA}}^2 & \rho\sigma_{\Delta\mathrm{RA}}\sigma_{\Delta\mathrm{DEC}} \\ \rho\sigma_{\Delta\mathrm{RA}}\sigma_{\Delta\mathrm{DEC}} & \sigma_{\Delta\mathrm{DEC}}^2\end{bmatrix} = \frac{1}{n_\mathrm{exp}}\mathrm{cov}\left\{\begin{pmatrix}\Delta\mathrm{RA} \\ \Delta\mathrm{DEC}\end{pmatrix}_k\right\}.
    \label{eq:covariance}
    \end{align}
\end{small}

\subsubsection{Off-axis swap observations}
\label{sec:astrometry_offaxis}
Contrary to the on-axis observations, extracting the metrology zero-point from the swap observations takes a bit more care. The overall method described in Section~\ref{sec:off-axis} holds true, and the metrology zero-point can be extracted by combining the arguments of the swapped and unswapped observations according to Equation~\ref{eq:zp_offaxis}. The main difficulty lies in the fact that all DITs are not acquired simultaneously, which causes some problems with how the individual DIT measurements are averaged, and how the two swap positions are combined. 

First, to avoid large errors on the phase, the coherent fluxes should be averaged before extracting the arguments. But due to the time-dependence of the baseline coordinates $u(t), v(t)$, this coherent flux average cannot be calculated simply by taking the average of all DITs, as this would "blur the signal". Instead, the proper way to average the coherent fluxes is to do it at zero-OPD, by first shifting them in the "target's reference frame", then calculating the average, and finally moving the result back to the initial frame. In mathematical terms, this would be written
\begin{equation}
    \begin{split}
    \Gamma^\mathrm{metcorr}_{j, k}(t_\mathrm{mean}, \lambda) = \left<\Gamma^\mathrm{metcorr}_{j, k}(t, \lambda)\times{}e^{i\frac{2\pi}{\lambda}\left[u(t)\Delta\mathrm{RA}+v(t)\Delta\mathrm{DEC}\right]}\right> \\ \times{}e^{-i\frac{2\pi}{\lambda}\left[u(t_\mathrm{mean})\Delta\mathrm{RA}+v(t_\mathrm{mean})\Delta\mathrm{DEC}\right]}.
    \end{split}
    \label{eq:proper_mean}
\end{equation}

Then, because the swapped and unswapped observations are not acquired simultaneously, Equation~\ref{eq:zp_offaxis} is not exactly correct, as it would in practice introduce an error on the phase in the form of $2\pi\Delta\mathrm{RA}\left[u(t_+) - u(t_-)\right]/\lambda$, plus a similar term for $\Delta\mathrm{DEC}$. Therefore, in practice, the zero-point is calculated directly in the zero-OPD frame using
\begin{equation}
    \begin{split}
    \phi_{j, k}^\mathrm{ZP} = & \frac{1}{2}\left<\Gamma^\mathrm{metcorr}_{j, k}e^{i\frac{2\pi}{\lambda}\left[u\Delta\mathrm{RA}+v\Delta\mathrm{DEC}\right]}\right>_\mathrm{swapped} \\ &+ \frac{1}{2}\left<\Gamma^\mathrm{metcorr}_{j, k}e^{i\frac{2\pi}{\lambda}\left[u\Delta\mathrm{RA}+v\Delta\mathrm{DEC}\right]}\right>_\mathrm{unswapped}.
    \end{split}
    \label{eq:proper_zp_offaxis}
\end{equation}

Of course, the problem in Equations~\ref{eq:proper_mean} and \ref{eq:proper_zp_offaxis} is that they require the a priori knowledge of the astrometry of the target. If this astrometry is known in advance, it can be used to average the swapped and unswapped observations, and then to calculate the metrology zero-point. This is typically the case when observing a known binary calibrator for an off-axis science target, but was not the case for our binary targets, for which we were actually trying to determine the best estimate of the astrometry. 

To circumvent this issue, we resorted to a grid of astrometry, in a similar fashion as for the on-axis observations described in Section~\ref{sec:on-axis}. For any point $\Delta{}\mathrm{RA}, \Delta\mathrm{DEC}$ on our grid, we combined the averaged coherent fluxes of the two swap positions using
\begin{small}
\begin{equation}
    \begin{split}
      \Gamma_{j, k}(\Delta\mathrm{RA}, \Delta\mathrm{DEC}) = & \left<\Gamma^\mathrm{metcorr}_{j, k}e^{i\frac{2\pi}{\lambda}\left[u\Delta\mathrm{RA}+v\Delta\mathrm{DEC}\right]}\right>_\mathrm{swapped}^\frac{1}{2} \\ &\times{}\mathrm{conj}\left[\left<\Gamma^\mathrm{metcorr}_{j, k}e^{i\frac{2\pi}{\lambda}\left[u\Delta\mathrm{RA}+v\Delta\mathrm{DEC}\right]}\right>_\mathrm{unswapped}\right]^\frac{1}{2}.
    \end{split}
\end{equation} 
\end{small}\noindent{}Thanks to the use of the complex-conjugate for the second element, the metrology zero-point vanishes. For the correct astrometry, this quantity should a purely real number. We therefore calculated an associated $\chi^2$ quantity from the imaginary part of this coherent flux:
\begin{equation}
    \chi^2(\Delta\mathrm{RA}, \Delta\mathrm{DEC}) = \sum_{(j, k), \lambda}\frac{\mathrm{Im}\left[\Gamma_{j, k}(\Delta\mathrm{RA}, \Delta\mathrm{DEC})(\lambda)\right]}{\sigma_{j, k}(\Delta\mathrm{RA}, \Delta\mathrm{DEC})(\lambda)}
\end{equation}
\noindent{}with $\sigma_{j, k}(\Delta\mathrm{RA}, \Delta\mathrm{DEC})$ the error bar on the imaginary part of $\Gamma$ propagated from the pipeline-reported error bars. The best astrometric guess was then taken as the location of the $\chi^2$ minimum:
\begin{equation}
    (\Delta\mathrm{RA}, \Delta\mathrm{DEC})_\mathrm{guess} = \mathrm{argmin}\left\{\chi^2(\Delta\mathrm{RA}, \Delta\mathrm{DEC})\right\}.
\end{equation}

The metrology zero-point was then calculated using Equation~\ref{eq:proper_zp_offaxis}, using this best guess to re-centre and average the coherent fluxes. Once the zero-point was known, we subtracted it from each individual exposure, and the rest of the data reduction followed the same steps as for the on-axis case, keeping in mind that the sign of all astrometric measurements from the swapped observations is reversed to account for the fact that these observations actually measure the position of the FT target relative to the SC target. This yielded one astrometric measurement per exposure file, and the final astrometric estimate was taken as the mean of the series, with the associated error bars calculated from the empirical covariance, using Equation~\ref{eq:covariance}. 

\begin{table*}
  \begin{center}
    \begin{tabular}{l l c c c c c c c c}
      \hline
      \hline
      Target & Alt. Name & RA & DEC & \multicolumn{3}{c}{$\chi^2_\mathrm{red}$ multiplicity test} & Contrast$^\dagger$ & Vetted \\
             &  & & & Epoch & Primary & Secondary & $\Delta\mathrm{mag}$ (K) & \\
      \hline
      WDSJ04021-3429 & HD 25535 & 04:02:03 & -34:28:56 & \multicolumn{3}{c}{\multirow{3}{*}{Vetted in ExoGRAVITY LP}} & 2.6 & YES \\
      WDSJ14077-4952 & HD 123227 & 14:07:42 & -49:52:03 & \multicolumn{3}{c}{} & 0.6 & YES \\
       WDSJ20399+1115 & HD 196885 & 20:39:52 & +11:14:59 & \multicolumn{3}{c}{} & 3.2 & YES \\
      WDSJ00209+1059 & HD 1663 & 00:20:55 & +10:58:38 & 2021-10-23 & 1.3 & 1.3 & 0.2 & YES \\
      WDSJ00427-3828 & $\lambda$ 01 Scl & 00:42:43 & -38:27:49 & 2021-10-21 & 1.3 & 1.3 & 0.3 &YES \\
      WDSJ05055+1948 & HD 32642 & 05:05:32 & +19:48:24 & 2021-11-03 & 1.2 & 1.2 & 0.3 & YES \\
      WDSJ05270-6837 & HD 36584 & 05:27:00 & -68:37:21 & 2020-12-01 & 1.8 & 1.5 & 0.3 & YES \\
      WDSJ06345-1114 & HD 46716 & 06:34:32 & -11:13:47 & 2021-10-22 & 1.2 & 1.3 & 0.2 & YES \\
      WDSJ06364+2717 & HD 46780 & 06:36:26 & +27:16:42 & 2020-12-17 & 1.6 & 1.6 & 0.7 & YES \\
      WDSJ08394-3636 & HD 73900 & 08:39:22 & -36:36:24 & 2021-10-25 & 1.2 & 1.2 & 1.4 & YES \\
      WDSJ10269+1713 & HD 90444 & 10:26:53 & +17:13:10 & 2022-02-06 & 1.4 & 1.3 & 0.3 & YES \\
      WDSJ10361-2641 & HD 91881 & 10:36:05 & -26:40:32 & 2020-02-06 & 1.5 & 1.4 & 1.0 & YES \\
      WDSJ18516-1719 & HD 174536 & 18:51:37 & -17:18:36 & 2023-03-30 & 1.6 & 1.7 & 3.5 & YES \\
      \hline
      \textcolor{gray}{WDSJ02407+2637} & \textcolor{gray}{HD 16638} & \textcolor{gray}{02:40:42} & \textcolor{gray}{+26:37:20} & \textcolor{gray}{2021-11-17} & \textcolor{gray}{63.1} & \textcolor{gray}{1.3} & \textcolor{gray}{-} & \textcolor{gray}{NO} \\      
      \textcolor{gray}{WDSJ05417-0254} & \textcolor{gray}{HD 37904} & \textcolor{gray}{05:41:40} & \textcolor{gray}{-02:53:47} & \textcolor{gray}{2020-12-01} & \textcolor{gray}{2.0} & \textcolor{gray}{7.8} & \textcolor{gray}{-} & \textcolor{gray}{NO} \\      
      \hline
    \end{tabular}
    \caption{Results of the $\chi^2_\mathrm{red}$ multiplicity test performed on the closure-phase for the first epoch of each target part of the calibration programme. This test shows that both WDSJ02407+2637 and WDSJ05417-0254 (in gray) have higher multiplicities, making them unsuitable as binary calibrators.\\
    $^\dagger$ For convenience, we also report An estimate of the contrast ratio for these targets, derived from our fits. Our observations were not designed for measuring the contrast, so these values should only be seen as rough estimates, and we do not report any error bars.}
    \label{tab:vetted}
  \end{center} 
\end{table*}

\section{Results}
\label{sec:results}

\subsection{Vetted selection of binary calibrators}

Among the 21 targets parts of the AT calibration programme to find new potential binary calibrators, 12 targets were observed (see Table~\ref{tab:selection}). Our $\chi^2$ closure-phase tests excluded WDSJ02407+2637 and WDSJ05417-0254, with respective maximum $\chi^2_\mathrm{red}$ values of 63 and 7.8 among their two visual components. These values are clear signs of a higher multiplicity, which justify their exclusion from the final list.

Overall, the calibration programme provided a set of 10 new binary calibrators. With the addition of the 3 calibrators already known at the start of the ExoGRAVITY Large Programme, this makes a total of 13 available binary calibrators for dual-field/off-axis observations with VLTI/GRAVITY on the UTs. The complete list of vetted binary calibrators, with coordinates and magnitudes, is provided in Table~\ref{tab:vetted}.

\begin{figure*}
    \centering
    \includegraphics[width=\linewidth, trim=0cm 0cm 0cm 2cm, clip]{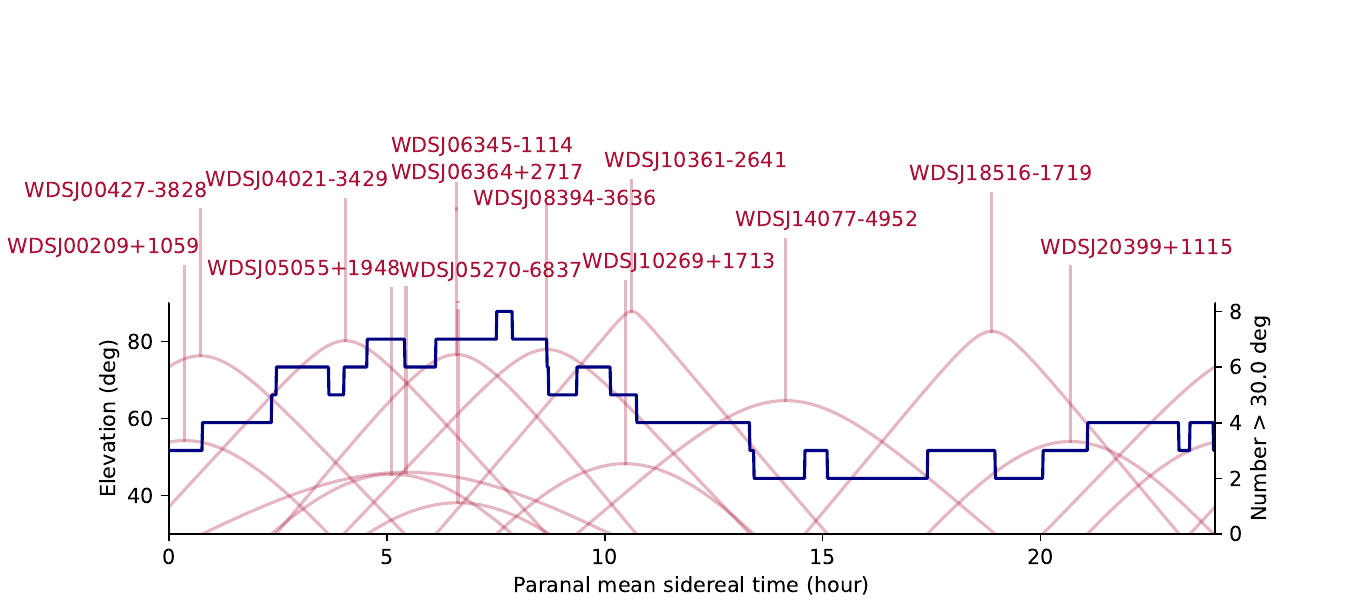}
    \caption{Observability of the dual-field calibrators as a function of mean sidereal time at Paranal. The red curves (left axis) show the elevation for each vetted binary calibrator from Table~\ref{tab:vetted}. The blue curve (right axis) shows the number of calibrators at above 30~deg of elevation.}
    \label{fig:elevation}
\end{figure*}

In Figure~\ref{fig:elevation}, we give the evolution of the elevation of these vetted calibrators as a function of sidereal time, for a complete rotation of the Earth (24h). We also give the evolution of the number of calibrators at an elevation of at least 30~deg, which shows that our list always guarantee the availability of at least 2 calibrators at an elevation $> 30\,\mathrm{deg}$ (airmass $< 2$), at any point in time. This should guarantee the feasibility of all upcoming programmes using the dual-field mode of GRAVITY on the UTs.

\subsection{Astrometry}

The astrometric measurements obtained on the vetted binaries with the ATs and UTs are presented in Table~\ref{tab:astrometry}. To obtain the best possible constraints on the orbits of our binaries, and to be able to provide estimates of the separation vector for the next few years, we added to our GRAVITY astrometric measurements a series of measurements taken from the WDS catalogue.

Our vetted binaries are all present in the catalogue, which provides astrometric measurements obtained with a mixture of instruments, some dating back to the 19th century. The catalogue does not provide the error bars on the individual measurements, but does provide the best estimate of the orbit of the binary obtained from the individual astrometric measurements. 

We used this best estimate of the orbit to extract an estimate of the error bars on the individual WDS data points. To do so, we first calculated the WDS astrometric residuals for all our vetted binaries, which we defined as the difference between each individual astrometric measurement available in the catalogue and position along the best estimate of the orbit at the corresponding epoch. Combining our 13 targets, we obtained a total of 1078 residuals in RA and DEC, corresponding to epochs ranging from 1843 to late 2021. We then adopted a simple linear model to describe the error bars, given by
\begin{equation}
\sigma(t) = b - at,
\end{equation}
\noindent{}in which $a$ was a positive coefficient describing the improvement in astrometric precision made over the years. We assumed that a residual $\delta\mathrm{RA}(t)$ or $\delta\mathrm{DEC}(t)$ obtained at epoch $t$ in the catalogue followed a normal distribution of mean 0 (unbiased), and variance $\sigma(t)^2$.

From there, we combined all the 1078 epochs gathered from the WDS catalogue to calculate a likelihood associated to our model of $\sigma$:
\begin{equation}
    \mathcal{L}\left[\sigma(t)\right] = \prod_t \frac{1}{2\pi\sigma(t)^2}\exp\left(-\frac{\delta{}\mathrm{RA}^2+\delta{}\mathrm{DEC}(t)^2}{2\sigma(t)^2}\right),
\end{equation}
\noindent{}and, ignoring the $2\pi$ factors in the normalization, we calculated the log-likelihood of the model using
\begin{equation}
    \log\mathcal{L}(a, b) = - \sum_t 2\log|b-at| - \frac{\delta\mathrm{RA}(t)^2 + \delta{}\mathrm{DEC}(t)^2}{2(b-at)^2}.
\end{equation}
\noindent{}We maximized this log-likelihood to obtain the best estimates of the parameters $a$ and $b$, which in turn provided our linear rule to calculate a value of $\sigma$ for all the individual measurements. We obtained:
\begin{equation}
    \frac{\sigma(t)}{1\,\mathrm{mas}} = 148 - 0.68 \times{} \left(\frac{t}{1\,\mathrm{yr}}-1836\right)
    \label{eq:sigma}
.\end{equation}

\subsection{Orbit fit and astrometric predictions}

\begin{table*}
\begin{center}
  \begin{tabular}{l c c c c c c}
  \hline
  \hline
  Target & \multicolumn{3}{c}{WDS catalogue} & \multicolumn{3}{c}{VLTI/GRAVITY (this work)} \\
         & $N_\mathrm{epochs}$ & First epoch & Last epoch & $N_\mathrm{epochs}$ & First epoch & Last epoch \\
    \hline
    WDSJ00209+1059 & 110 & 1889-08-26 & 2016-11-16 & 1 & 2021-10-23 & 2021-10-23\\
    WDSJ00427-3828 & 42 & 1901-11-03 & 2017-10-17 & 3 & 2021-10-21 & 2021-11-16\\
    WDSJ04021-3429 & 95 & 1881-11-07 & 2021-11-21 & 11 & 2019-10-21 & 2023-11-02\\
    WDSJ05055+1948 & 203 & 1843-03-25 & 2021-02-26 & 3 & 2021-11-03 & 2021-12-28\\
    WDSJ05270-6837 & 38 & 1898-01-01 & 2015-01-01 & 3 & 2020-12-01 & 2022-09-15\\
    WDSJ06345-1114 & 28 & 1888-08-22 & 2016-03-16 & 2 & 2021-10-22 & 2021-12-28\\
    WDSJ06364+2717 & 165 & 1843-03-22 & 2021-02-26 & 1 & 2020-12-17 & 2020-12-17\\
    WDSJ08394-3636 & 33 & 1900-05-19 & 2021-11-21 & 5 & 2021-10-25 & 2022-01-24\\
    WDSJ10269+1713 & 166 & 1843-04-05 & 2019-02-09 & 2 & 2022-02-26 & 2022-02-28\\
    WDSJ10361-2641 & 113 & 1876-03-25 & 2021-04-26 & 8 & 2020-02-06 & 2023-05-10\\
    WDSJ14077-4952 & 59 & 1895-06-25 & 2021-07-26 & 5 & 2020-02-07 & 2023-06-02\\
    WDSJ18516-1719 & 13 & 1900-09-06 & 2002-08-05 & 4 & 2021-08-27 & 2023-07-02\\
    WDSJ20399+1115 & 13 & 2000-10-08 & 2021-10-02 & 5 & 2019-08-15 & 2021-10-20\\
    \hline
  \end{tabular}  
  \caption{Summary of the astrometric measurements used in our orbit fits.}
  \label{tab:measurements}
  \end{center}
\end{table*}

\begin{table*}
  \begin{center}
  \begin{tabular}{l | c c c | c c c | c c c }
    \hline
    \hline
Target & \multicolumn{3}{c}{2024-01-01} & \multicolumn{3}{c}{2025-01-01} & \multicolumn{3}{c}{2026-01-01} \\
 & $\Delta{}\mathrm{RA}$ & $\Delta{}\mathrm{DEC}$ & Error$^1$ & $\Delta{}\mathrm{RA}$ & $\Delta{}\mathrm{DEC}$ & Error$^1$ & $\Delta{}\mathrm{RA}$ & $\Delta{}\mathrm{DEC}$ & Error$^1$ \\
 & [mas] & [mas] & [mas] & [mas] & [mas] & [mas] & [mas] & [mas] & [mas] \\
\hline
WDSJ00209+1059 & $682.7$ & $-405.9$ & $0.9$  & $682.5$ & $-409.7$ & $1.1$  & $682.3$ & $-413.3$ & $1.5$ \\
WDSJ00427-3828 & $-297.5$ & $-615.1$ & $1.2$  & $-300.6$ & $-610.4$ & $1.8$  & $-303.6$ & $-605.7$ & $2.4$ \\
WDSJ04021-3429 & $718.7$ & $802.8$ & $0.3$  & $691.7$ & $814.5$ & $0.4$  & $664.3$ & $825.5$ & $0.7$ \\
WDSJ05055+1948 & $-865.2$ & $378.2$ & $0.5$  & $-867.0$ & $374.4$ & $0.7$  & $-868.7$ & $370.5$ & $0.9$ \\
WDSJ05270-6837 & $553.5$ & $-1353.8$ & $0.7$  & $558.1$ & $-1354.8$ & $1.0$  & $562.7$ & $-1355.8$ & $1.4$ \\
WDSJ06345-1114 & $131.5$ & $708.8$ & $1.7$  & $138.8$ & $714.7$ & $2.5$  & $146.1$ & $720.5$ & $3.3$ \\
WDSJ06364+2717 & $-694.0$ & $-4.3$ & $2.9$  & $-683.8$ & $-18.7$ & $4.0$  & $-671.4$ & $-33.0$ & $5.4$ \\
WDSJ08394-3636 & $-833.0$ & $-468.6$ & $2.2$  & $-834.4$ & $-473.9$ & $3.3$  & $-834.5$ & $-478.4$ & $4.4$ \\
WDSJ10269+1713 & $419.6$ & $-722.2$ & $0.6$  & $419.7$ & $-725.5$ & $0.8$  & $419.7$ & $-728.6$ & $1.1$ \\
WDSJ10361-2641 & $-1114.0$ & $681.3$ & $0.4$  & $-1115.2$ & $671.2$ & $0.6$  & $-1116.1$ & $660.9$ & $0.7$ \\
WDSJ14077-4952 & $-385.8$ & $884.8$ & $0.3$  & $-362.3$ & $878.6$ & $0.5$  & $-338.7$ & $871.9$ & $0.7$ \\
WDSJ18516-1719 & $-314.5$ & $1486.4$ & $0.3$  & $-314.4$ & $1486.4$ & $0.4$  & $-314.2$ & $1486.4$ & $0.5$ \\
WDSJ20399+1115 & $231.4$ & $423.7$ & $2.9$  & $195.2$ & $418.7$ & $5.3$  & $158.2$ & $411.8$ & $8.7$ \\
\hline
  \end{tabular}
  \caption{Predicted astrometry from 2024 to 2026 for the vetted binary calibrators. \\
  $ ^1$ The astrometric predictions are calculated form the posteriors of the orbit fits, and the RA and DEC can be highly correlated. Consequently, the error reported in the tables represent the long-axis of the error ellipse.}
  \label{tab:predictions}
  \end{center}  
\end{table*}

We used \verb|orbitize!| \citep{Blunt2020} to fit the relative motion of our vetted binaries. For each of them, we gathered all the astrometric measurements available in the WDS catalogue, to which we associated error bars using Equation~\ref{eq:sigma}. We also added our own GRAVITY astrometric measurements, as reported in Table~\ref{tab:astrometry}. Table~\ref{tab:measurements} gives a brief overview of the measurements available on each target.

We used the parallel-tempering affine-invariant \citep{Foreman-Mackey2013, Vousden2016} MCMC sampler available in \verb|orbitize!| to determine the posterior distributions of the orbital parameters. Except for the total mass of the system $M_\mathrm{tot}$ and the parallax $\pi$, we used the default priors in \verb|orbitize!|. Our priors on the total mass were set to Gaussian priors, with an arbitrary $M_\mathrm{tot}=4\pm{}4$. For the parallax, we based our priors on the Gaia DR3 measurements \cite{GaiaCollaboration2021}. For systems which remained unresolved with Gaia, and for which only one value of the parallax was available, we adopted a Gaussian prior centred on the Gaia measurement, and with a standard-deviation corresponding to the reported measurement error. For systems on which measurements were available for each component, we took the average of the two as our mean value for the Gaussian prior, and we used the half difference of the two components as our prior standard deviation.

\begin{figure*}
  \begin{center}
    \includegraphics[width=0.95\linewidth]{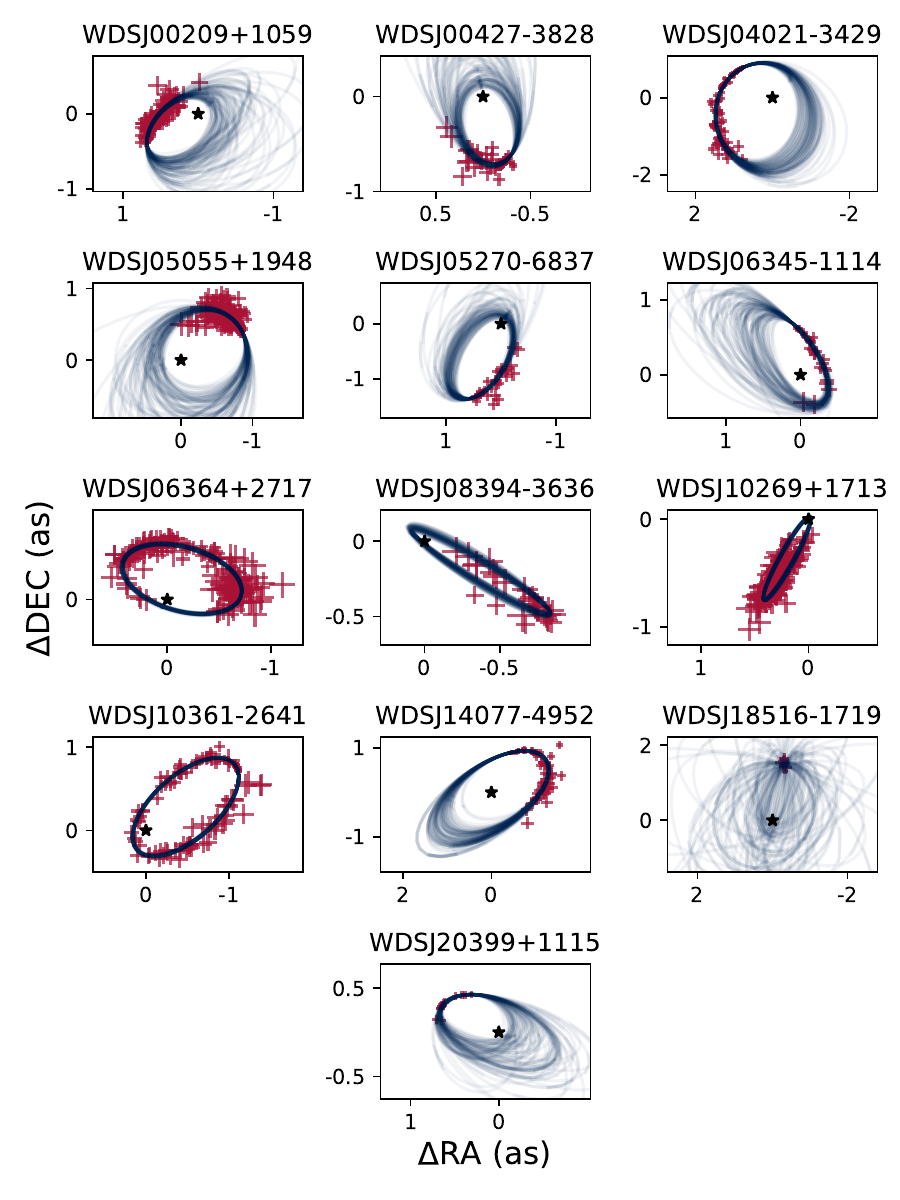}
    \label{fig:orbits}    
    \caption{Posterior orbits for each of the 13 vetted calibrators from Table~\ref{tab:vetted}. Each panel illustrates a set of 100 orbits drawn from our posterior distributions of orbital parameters (in blue), as well as the coverage offered by our combined WDS and GRAVITY astrometry (red crosses). The black star provides a visual reference of the $(0, 0)$ coordinate origin.}
  \end{center}
\end{figure*}

The MCMC runs were all set up with 100 walkers and 20 temperatures for a burn-in phase of $300\,000$ steps, followed by an additional $600\,000$ steps to approximate the posterior distributions, keeping only 1/10th of the samples. The resulting orbital parameter estimates are reported in Table~\ref{tab:parameters}, and Figure~\ref{fig:orbits} gives a visual overview of these posteriors for each target. In particular, Figure~\ref{fig:orbits} clearly illustrate the very large disparities in terms of orbital coverage, which inevitably results in very large disparities in terms of orbital constraints.

From these orbital posteriors, we also calculated a set of astrometric predictions which can be used for future observations. To obtain these predictions, we calculated the predicted astrometry for each orbit of our posteriors at the beginning of 2024, 2025, and 2026. We report the mean values and the error estimates in  Table~\ref{tab:predictions}, but we strongly encourage the use of the tool \texttt{whereistheplanet}\footnote{Available at https://github.com/semaphoreP/whereistheplanet} to calculate predicted separations, as we plan to update the orbit posteriors to improve predictions when new observations become available. This tool also seamlessly integrates with \texttt{p2Gravity}\footnote{Available at https://github.com/f4hzg/p2Gravity}, to further simplify the process of preparing observing material for VLTI/GRAVITY. All the vetted calibrators of Table~\ref{tab:vetted} have been integrated in \texttt{whereistheplanet}.

\section{Conclusion}
\label{sec:conclusion}

We have assembled a catalogue of 13 visual binary stars which can be used as calibrators for dual-field interferometric observations with VLTI/GRAVITY on the UTs. This catalogue guarantees that there are always at least 2 vetted calibrators up in the sky above Paranal.

Using a combination of our own recent high-precision astrometric measurements together with archival measurements from the WDS catalogue, we also obtained updated orbital elements for the 13 binaries of the catalogue. From these updated orbital elements, we derive a set of predictions for the binary separation vector which can be used to point the instrument during calibration observations. These predictions are available in Table~\ref{tab:predictions} and in \texttt{whereistheplanet}, and we recommend the use of this tool together with \texttt{p2Gravity} for the preparation of VLTI/GRAVITY dual-field observing material.

\begin{acknowledgements}
This work is based on observations collected at the European Southern Observatory under ESO programmes 1103.B-0626, 1104.C-0651, 106.21S5.001, 108.226K.002, 108.22MY.001, 112.25GE.001, 111.24MN.001, 2104.C-5040. 5102.C-0122.    
SL acknowledges the support of the French Agence Nationale de la Recherche (ANR), under grant ANR-21-CE31-0017 (project ExoVLTI).
G-DM acknowledges the support of the DFG priority program SPP 1992 ``Exploring the Diversity of Extrasolar Planets'' (MA~9185/1) and from the Swiss National Science Foundation under grant 200021\_204847 ``PlanetsInTime''. Parts of this work have been carried out within the framework of the NCCR PlanetS supported by the Swiss National Science Foundation.

\end{acknowledgements}

\bibliographystyle{aa} 
\bibliography{newbib} 

\begin{thebibliography}{18}
\expandafter\ifx\csname natexlab\endcsname\relax\def\natexlab#1{#1}\fi

\bibitem[{Blunt {et~al.}(2020)Blunt, Wang, Angelo, Ngo, Cody, Rosa, Graham,
  Hirsch, Nagpal, Nielsen, Pearce, Rice, \& Tejada}]{Blunt2020}
Blunt, S., Wang, J.~J., Angelo, I., {et~al.} 2020, The Astronomical Journal,
  159, 89

\bibitem[{{Foreman-Mackey} {et~al.}(2013){Foreman-Mackey}, Hogg, Lang, \&
  Goodman}]{Foreman-Mackey2013}
{Foreman-Mackey}, D., Hogg, D.~W., Lang, D., \& Goodman, J. 2013, Publications
  of the Astronomical Society of the Pacific, 125, 306

\bibitem[{{Gaia Collaboration} {et~al.}(2021){Gaia Collaboration}, Brown,
  Vallenari, Prusti, de~Bruijne, Babusiaux, Biermann, Creevey, Evans, Eyer,
  Hutton, Jansen, Jordi, Klioner, Lammers, Lindegren, Luri, Mignard, Panem,
  Pourbaix, Randich, Sartoretti, Soubiran, Walton, Arenou, {Bailer-Jones},
  Bastian, Cropper, Drimmel, Katz, Lattanzi, van Leeuwen, Bakker, Cacciari,
  Casta{\~n}eda, Angeli, Ducourant, Fabricius, Fouesneau, Fr{\'e}mat, Guerra,
  Guerrier, Guiraud, Piccolo, Masana, Messineo, Mowlavi, Nicolas, Nienartowicz,
  Pailler, Panuzzo, Riclet, Roux, Seabroke, Sordo, Tanga, Th{\'e}venin,
  {Gracia-Abril}, Portell, Teyssier, Altmann, Andrae, {Bellas-Velidis}, Benson,
  Berthier, Blomme, Brugaletta, Burgess, Busso, Carry, Cellino, Cheek,
  Clementini, Damerdji, Davidson, Delchambre, Dell'Oro,
  {Fern{\'a}ndez-Hern{\'a}ndez}, Galluccio, {Garc{\'i}a-Lario},
  {Garcia-Reinaldos}, {Gonz{\'a}lez-N{\'u}{\~n}ez}, Gosset, Haigron, Halbwachs,
  Hambly, Harrison, Hatzidimitriou, Heiter, Hern{\'a}ndez, Hestroffer, Hodgkin,
  Holl, Jan{\ss}en, de~Fombelle, Jordan, {Krone-Martins}, Lanzafame,
  L{\"o}ffler, Lorca, Manteiga, Marchal, Marrese, Moitinho, Mora, Muinonen,
  Osborne, Pancino, Pauwels, Petit, {Recio-Blanco}, Richards, Riello,
  Rimoldini, Robin, Roegiers, Rybizki, Sarro, Siopis, Smith, Sozzetti, Ulla,
  Utrilla, van Leeuwen, van Reeven, Abbas, Aramburu, Accart, Aerts, Aguado,
  Ajaj, Altavilla, {\'A}lvarez, {Cid-Fuentes}, Alves, Anderson, Varela, Antoja,
  Audard, Baines, Baker, {Balaguer-N{\'u}{\~n}ez}, Balbinot, Balog, Barache,
  Barbato, Barros, Barstow, Bartolom{\'e}, Bassilana, Bauchet,
  {Baudesson-Stella}, Becciani, Bellazzini, Bernet, Bertone, Bianchi,
  {Blanco-Cuaresma}, Boch, Bombrun, Bossini, Bouquillon, Bragaglia, Bramante,
  Breedt, Bressan, Brouillet, Bucciarelli, Burlacu, Busonero, Butkevich, Buzzi,
  Caffau, Cancelliere, C{\'a}novas, {Cantat-Gaudin}, Carballo, Carlucci,
  Carnerero, Carrasco, Casamiquela, Castellani, {Castro-Ginard}, Sampol,
  Chaoul, Charlot, Chemin, Chiavassa, Cioni, Comoretto, Cooper, Cornez, Cowell,
  Crifo, Crosta, Crowley, Dafonte, Dapergolas, David, David, de~Laverny, Luise,
  March, Ridder, de~Souza, de~Teodoro, de~Torres, del Peloso, del Pozo, Delbo,
  Delgado, Delgado, Delisle, Matteo, Diakite, Diener, Distefano, Dolding,
  Eappachen, Edvardsson, Enke, Esquej, Fabre, Fabrizio, Faigler, Fedorets,
  Fernique, Fienga, Figueras, Fouron, Fragkoudi, Fraile, Franke, Gai, Garabato,
  {Garcia-Gutierrez}, {Garc{\'i}a-Torres}, Garofalo, Gavras, Gerlach, Geyer,
  Giacobbe, Gilmore, Girona, Giuffrida, Gomel, Gomez, {Gonzalez-Santamaria},
  {Gonz{\'a}lez-Vidal}, Granvik, {Guti{\'e}rrez-S{\'a}nchez}, Guy, Hauser,
  Haywood, Helmi, Hidalgo, Hilger, H{\l}adczuk, Hobbs, Holland, Huckle,
  Jasniewicz, Jonker, Campillo, Julbe, Karbevska, Kervella, Khanna, Kochoska,
  Kontizas, Kordopatis, Korn, {Kostrzewa-Rutkowska}, Kruszy{\'n}ska, Lambert,
  Lanza, Lasne, Campion, Fustec, Lebreton, Lebzelter, Leccia, Leclerc,
  {Lecoeur-Taibi}, Liao, Licata, Lindstr{\o}m, Lister, Livanou, Lobel, Pardo,
  Managau, Mann, Marchant, Marconi, Santos, Marinoni, Marocco, Marshall, Polo,
  {Mart{\'i}n-Fleitas}, Masip, Massari, {Mastrobuono-Battisti}, Mazeh,
  McMillan, Messina, Michalik, Millar, Mints, Molina, Molinaro, Moln{\'a}r,
  Montegriffo, Mor, Morbidelli, Morel, Morris, Mulone, Munoz, Muraveva, Murphy,
  Musella, Noval, Ord{\'e}novic, Orr{\`u}, Osinde, Pagani, Pagano, Palaversa,
  Palicio, Panahi, Pawlak, Esteller, Penttil{\"a}, Piersimoni, Pineau, Plachy,
  Plum, Poggio, Poretti, Poujoulet, Pr{\v s}a, Pulone, Racero, Ragaini, Rainer,
  Raiteri, Rambaux, Ramos, {Ramos-Lerate}, Fiorentin, Regibo, Reyl{\'e},
  Ripepi, Riva, Rixon, Robichon, Robin, Roelens, Rohrbasser,
  {Romero-G{\'o}mez}, Rowell, Royer, Rybicki, Sadowski, Sell{\'e}s, Sahlmann,
  Salgado, Salguero, Samaras, Gimenez, Sanna, Santove{\~n}a, Sarasso,
  Schultheis, Sciacca, Segol, Segovia, S{\'e}gransan, Semeux, Shahaf, Siddiqui,
  Siebert, Siltala, Slezak, Smart, Solano, Solitro, Souami, Souchay, Spagna,
  Spoto, Steele, Steidelm{\"u}ller, Stephenson, S{\"u}veges, Szabados,
  {Szegedi-Elek}, Taris, Tauran, Taylor, Teixeira, Thuillot, Tonello, Torra,
  Torra, Turon, Unger, Vaillant, van Dillen, Vanel, Vecchiato, Viala, Vicente,
  Voutsinas, Weiler, Wevers, Wyrzykowski, Yoldas, Yvard, Zhao, Zorec, Zucker,
  Zurbach, \& Zwitter}]{GaiaCollaboration2021}
{Gaia Collaboration}, Brown, A. G.~A., Vallenari, A., {et~al.} 2021, Astronomy
  \& Astrophysics, 649, A1

\bibitem[{Gillessen {et~al.}(2012)Gillessen, Lippa, Eisenhauer, Pfuhl, Haug,
  Kellner, Ott, Wieprecht, Sturm, Hau{\ss}mann, Kister, Moch, \&
  Thiel}]{Gillessen2012}
Gillessen, S., Lippa, M., Eisenhauer, F., {et~al.} 2012, in Optical and
  {{Infrared Interferometry III}}, Vol. 8445 ({International Society for Optics
  and Photonics}), 84451O

\bibitem[{{GRAVITY Collaboration} {et~al.}(2017){GRAVITY Collaboration},
  Abuter, Accardo, Amorim, Anugu, {\'A}vila, Azouaoui, Benisty, Berger, Blind,
  Bonnet, Bourget, Brandner, Brast, Buron, Burtscher, Cassaing, Chapron,
  Choquet, Cl{\'e}net, Collin, du~Foresto, de~Wit, de~Zeeuw, Deen,
  {Delplancke-Str{\"o}bele}, Dembet, Derie, Dexter, Duvert, Ebert, Eckart,
  Eisenhauer, Esselborn, F{\'e}dou, Finger, Garcia, Dabo, Lopez, Gendron,
  Genzel, Gillessen, Gonte, Gordo, Grould, Gr{\"o}zinger, Guieu, Haguenauer,
  Hans, Haubois, Haug, Haussmann, Henning, Hippler, Horrobin, Huber, Hubert,
  Hubin, Hummel, Jakob, Janssen, Jochum, Jocou, Kaufer, Kellner, Kendrew, Kern,
  Kervella, Kiekebusch, Klein, Kok, Kolb, Kulas, Lacour, Lapeyr{\`e}re,
  Lazareff, Bouquin, L{\`e}na, Lenzen, L{\'e}v{\^e}que, Lippa, Magnard,
  Mehrgan, Mellein, M{\'e}rand, {Moreno-Ventas}, Moulin, M{\"u}ller,
  M{\"u}ller, Neumann, Oberti, Ott, Pallanca, Panduro, Pasquini, Paumard,
  Percheron, Perraut, Perrin, Pfl{\"u}ger, Pfuhl, Duc, Plewa, Popovic, Rabien,
  Ram{\'i}rez, Ramos, Rau, Riquelme, Rohloff, Rousset, {Sanchez-Bermudez},
  Scheithauer, Sch{\"o}ller, Schuhler, Spyromilio, Straubmeier, Sturm, Suarez,
  Tristram, Ventura, Vincent, Waisberg, Wank, Weber, Wieprecht, Wiest,
  Wiezorrek, Wittkowski, Woillez, Wolff, Yazici, Ziegler, \&
  Zins}]{GRAVITYCollaboration2017}
{GRAVITY Collaboration}, Abuter, R., Accardo, M., {et~al.} 2017, Astronomy \&
  Astrophysics, 602, A94

\bibitem[{{GRAVITY Collaboration} {et~al.}(2022){GRAVITY Collaboration},
  Abuter, Aimar, Amorim, Ball, Baub{\"o}ck, Berger, Bonnet, Bourdarot,
  Brandner, Cardoso, Cl{\'e}net, Dallilar, Davies, de~Zeeuw, Dexter, Drescher,
  Eisenhauer, Schreiber, Foschi, Garcia, Gao, Gendron, Genzel, Gillessen,
  Habibi, Haubois, Hei{\ss}el, Henning, Hippler, Horrobin, Jochum, Jocou,
  Kaufer, Kervella, Lacour, Lapeyr{\`e}re, Bouquin, L{\'e}na, Lutz, Ott,
  Paumard, Perraut, Perrin, Pfuhl, Rabien, Shangguan, Shimizu, Scheithauer,
  Stadler, Stephens, Straub, Straubmeier, Sturm, Tacconi, Tristram, Vincent,
  von Fellenberg, Widmann, Wieprecht, Wiezorrek, Woillez, Yazici, \&
  Young}]{GRAVITYCollaboration2022}
{GRAVITY Collaboration}, Abuter, R., Aimar, N., {et~al.} 2022, Astronomy \&
  Astrophysics, 657, L12

\bibitem[{{GRAVITY Collaboration} {et~al.}(2018{\natexlab{a}}){GRAVITY
  Collaboration}, Abuter, Amorim, Anugu, Baub{\"o}ck, Benisty, Berger, Blind,
  Bonnet, Brandner, Buron, Collin, Chapron, Cl{\'e}net, dCoud{\'e}~u Foresto,
  de~Zeeuw, Deen, {Delplancke-Str{\"o}bele}, Dembet, Dexter, Duvert, Eckart,
  Eisenhauer, Finger, Schreiber, F{\'e}dou, Garcia, Lopez, Gao, Gendron,
  Genzel, Gillessen, Gordo, Habibi, Haubois, Haug, Hau{\ss}mann, Henning,
  Hippler, Horrobin, Hubert, Hubin, Rosales, Jochum, Jocou, Kaufer, Kellner,
  Kendrew, Kervella, Kok, Kulas, Lacour, Lapeyr{\`e}re, Lazareff, Bouquin,
  L{\'e}na, Lippa, Lenzen, M{\'e}rand, M{\"u}ler, Neumann, Ott, Palanca,
  Paumard, Pasquini, Perraut, Perrin, Pfuhl, Plewa, Rabien, Ram{\'i}rez, Ramos,
  Rau, {Rodr{\'i}guez-Coira}, Rohloff, Rousset, {Sanchez-Bermudez},
  Scheithauer, Sch{\"o}ller, Schuler, Spyromilio, Straub, Straubmeier, Sturm,
  Tacconi, Tristram, Vincent, von Fellenberg, Wank, Waisberg, Widmann,
  Wieprecht, Wiest, Wiezorrek, Woillez, Yazici, Ziegler, \&
  Zins}]{GRAVITYCollaboration2018a}
{GRAVITY Collaboration}, Abuter, R., Amorim, A., {et~al.} 2018{\natexlab{a}},
  Astronomy \& Astrophysics, 615, L15

\bibitem[{{GRAVITY Collaboration} {et~al.}(2020{\natexlab{a}}){GRAVITY
  Collaboration}, Abuter, Amorim, Baub{\"o}ck, Berger, Bonnet, Brandner,
  Cardoso, Cl{\'e}net, de~Zeeuw, Dexter, Eckart, Eisenhauer, Schreiber, Garcia,
  Gao, Gendron, Genzel, Gillessen, Habibi, Haubois, Henning, Hippler, Horrobin,
  {Jim{\'e}nez-Rosales}, Jochum, Jocou, Kaufer, Kervella, Lacour,
  Lapeyr{\`e}re, Bouquin, L{\'e}na, Nowak, Ott, Paumard, Perraut, Perrin,
  Pfuhl, {Rodr{\'i}guez-Coira}, Shangguan, Scheithauer, Stadler, Straub,
  Straubmeier, Sturm, Tacconi, Vincent, von Fellenberg, Waisberg, Widmann,
  Wieprecht, Wiezorrek, Woillez, Yazici, \& Zins}]{GRAVITYCollaboration2020c}
{GRAVITY Collaboration}, Abuter, R., Amorim, A., {et~al.} 2020{\natexlab{a}},
  Astronomy \& Astrophysics, 636, L5

\bibitem[{{GRAVITY Collaboration} {et~al.}(2018{\natexlab{b}}){GRAVITY
  Collaboration}, Abuter, Amorim, Baub{\"o}ck, Berger, Bonnet, Brandner,
  Cl{\'e}net, du~Foresto, de~Zeeuw, Deen, Dexter, Duvert, Eckart, Eisenhauer,
  Schreiber, Garcia, Gao, Gendron, Genzel, Gillessen, Guajardo, Habibi,
  Haubois, Henning, Hippler, Horrobin, Huber, {Jim{\'e}nez-Rosales}, Jocou,
  Kervella, Lacour, Lapeyr{\`e}re, Lazareff, Bouquin, L{\'e}na, Lippa, Ott,
  Panduro, Paumard, Perraut, Perrin, Pfuhl, Plewa, Rabien,
  {Rodr{\'i}guez-Coira}, Rousset, Sternberg, Straub, Straubmeier, Sturm,
  Tacconi, Vincent, von Fellenberg, Waisberg, Widmann, Wieprecht, Wiezorrek,
  Woillez, \& Yazici}]{GRAVITYCollaboration2018b}
{GRAVITY Collaboration}, Abuter, R., Amorim, A., {et~al.} 2018{\natexlab{b}},
  Astronomy \& Astrophysics, 618, L10

\bibitem[{{GRAVITY Collaboration} {et~al.}(2019{\natexlab{a}}){GRAVITY
  Collaboration}, Abuter, Amorim, Baub{\"o}ck, Berger, Bonnet, Brandner,
  Cl{\'e}net, du~Foresto, de~Zeeuw, Dexter, Duvert, Eckart, Eisenhauer,
  Schreiber, Garcia, Gao, Gendron, Genzel, Gerhard, Gillessen, Habibi, Haubois,
  Henning, Hippler, Horrobin, {Jim{\'e}nez-Rosales}, Jocou, Kervella, Lacour,
  Lapeyr{\`e}re, Bouquin, L{\'e}na, Ott, Paumard, Perraut, Perrin, Pfuhl,
  Rabien, Coira, Rousset, Scheithauer, Sternberg, Straub, Straubmeier, Sturm,
  Tacconi, Vincent, von Fellenberg, Waisberg, Widmann, Wieprecht, Wiezorrek,
  Woillez, \& Yazici}]{GRAVITYCollaboration2019b}
{GRAVITY Collaboration}, Abuter, R., Amorim, A., {et~al.} 2019{\natexlab{a}},
  Astronomy \& Astrophysics, 625, L10

\bibitem[{{GRAVITY Collaboration} {et~al.}(2020{\natexlab{b}}){GRAVITY
  Collaboration}, Baub{\"o}ck, Dexter, Abuter, Amorim, Berger, Bonnet,
  Brandner, Cl{\'e}net, du~Foresto, de~Zeeuw, Duvert, Eckart, Eisenhauer,
  Schreiber, Gao, Garcia, Gendron, Genzel, Gerhard, Gillessen, Habibi, Haubois,
  Henning, Hippler, Horrobin, {Jim{\'e}nez-Rosales}, Jocou, Kervella, Lacour,
  Lapeyr{\`e}re, Bouquin, L{\'e}na, Ott, Paumard, Perraut, Perrin, Pfuhl,
  Rabien, Coira, Rousset, Scheithauer, Stadler, Sternberg, Straub, Straubmeier,
  Sturm, Tacconi, Vincent, von Fellenberg, Waisberg, Widmann, Wieprecht,
  Wiezorrek, Woillez, \& Yazici}]{GRAVITYCollaboration2020b}
{GRAVITY Collaboration}, Baub{\"o}ck, M., Dexter, J., {et~al.}
  2020{\natexlab{b}}, Astronomy \& Astrophysics, 635, A143

\bibitem[{{GRAVITY Collaboration} {et~al.}(2019{\natexlab{b}}){GRAVITY
  Collaboration}, Lacour, Nowak, Wang, Pfuhl, Eisenhauer, Abuter, Amorim,
  Anugu, Benisty, Berger, Beust, Blind, Bonnefoy, Bonnet, Bourget, Brandner,
  Buron, Collin, Charnay, Chapron, Cl{\'e}net, Coud{\'e} Du~Foresto, {de
  Zeeuw}, Deen, Dembet, Dexter, Duvert, Eckart, F{\"o}rster~Schreiber,
  F{\'e}dou, Garcia, Garcia~Lopez, Gao, Gendron, Genzel, Gillessen, Gordo,
  Greenbaum, Habibi, Haubois, Hau{\ss}mann, Henning, Hippler, Horrobin, Hubert,
  Jimenez~Rosales, Jocou, Kendrew, Kervella, Kolb, Lagrange, Lapeyr{\`e}re,
  Le~Bouquin, L{\'e}na, Lippa, Lenzen, Maire, Molli{\`e}re, Ott, Paumard,
  Perraut, Perrin, Pueyo, Rabien, Ram{\'i}rez, Rau, {Rodr{\'i}guez-Coira},
  Rousset, {Sanchez-Bermudez}, Scheithauer, Schuhler, Straub, Straubmeier,
  Sturm, Tacconi, Vincent, {van Dishoeck}, {von Fellenberg}, Wank, Waisberg,
  Widmann, Wieprecht, Wiest, Wiezorrek, Woillez, Yazici, Ziegler, \&
  Zins}]{GRAVITYCollaboration2019a}
{GRAVITY Collaboration}, Lacour, S., Nowak, M., {et~al.} 2019{\natexlab{b}},
  Astronomy and Astrophysics, 623, L11

\bibitem[{{GRAVITY Collaboration} {et~al.}(2020{\natexlab{c}}){GRAVITY
  Collaboration}, Nowak, Lacour, Molli{\`e}re, Wang, Charnay, van Dishoeck,
  Abuter, Amorim, Berger, Beust, Bonnefoy, Bonnet, Brandner, Buron,
  Cantalloube, Collin, Chapron, Cl{\'e}net, du~Foresto, de~Zeeuw, Dembet,
  Dexter, Duvert, Eckart, Eisenhauer, Schreiber, F{\'e}dou, Lopez, Gao,
  Gendron, Genzel, Gillessen, Hau{\ss}mann, Henning, Hippler, Hubert, Jocou,
  Kervella, Lagrange, Lapeyr{\`e}re, Bouquin, L{\'e}na, Maire, Ott, Paumard,
  Paladini, Perraut, Perrin, Pueyo, Pfuhl, Rabien, Rau, {Rodr{\'i}guez-Coira},
  Rousset, Scheithauer, Shangguan, Straub, Straubmeier, Sturm, Tacconi,
  Vincent, Widmann, Wieprecht, Wiezorrek, Woillez, Yazici, \&
  Ziegler}]{GRAVITYCollaboration2020d}
{GRAVITY Collaboration}, Nowak, M., Lacour, S., {et~al.} 2020{\natexlab{c}},
  Astronomy \& Astrophysics, 633, A110

\bibitem[{{GRAVITY Collaboration} {et~al.}(2020{\natexlab{d}}){GRAVITY
  Collaboration}, Pfuhl, Davies, Dexter, Netzer, H{\"o}nig, Lutz, Schartmann,
  Sturm, Amorim, Brandner, Cl{\'e}net, de~Zeeuw, Eckart, Eisenhauer, Schreiber,
  Gao, Garcia, Genzel, Gillessen, Gratadour, Kishimoto, Lacour, Millour, Ott,
  Paumard, Perraut, Perrin, Peterson, Petrucci, Prieto, Rouan, Shangguan,
  Shimizu, Sternberg, Straub, Straubmeier, Tacconi, Tristram, Vermot, Waisberg,
  Widmann, \& Woillez}]{GRAVITYCollaboration2020}
{GRAVITY Collaboration}, Pfuhl, O., Davies, R., {et~al.} 2020{\natexlab{d}},
  Astronomy \& Astrophysics, 634, A1

\bibitem[{Lacour {et~al.}(2019)Lacour, Dembet, Abuter, F{\'e}dou, Perrin,
  Choquet, Pfuhl, Eisenhauer, Woillez, Cassaing, Wieprecht, Ott, Wiezorrek,
  Tristram, Wolff, Ram{\'i}rez, Haubois, Perraut, Straubmeier, Brandner, \&
  Amorim}]{Lacour2019}
Lacour, S., Dembet, R., Abuter, R., {et~al.} 2019, Astronomy \& Astrophysics,
  624, A99

\bibitem[{Lapeyrere {et~al.}(2014)Lapeyrere, Kervella, Lacour, Azouaoui,
  {Garcia-Dabo}, Perrin, Eisenhauer, Perraut, Straubmeier, Amorim, \&
  Brandner}]{Lapeyrere2014}
Lapeyrere, V., Kervella, P., Lacour, S., {et~al.} 2014, in Optical and
  {{Infrared Interferometry IV}}, Vol. 9146 ({SPIE}), 735--743

\bibitem[{Nowak {et~al.}(2020)Nowak, Lacour, Lagrange, Rubini, Wang, Stolker,
  Abuter, Amorim, {Asensio-Torres}, Baub{\"o}ck, Benisty, Berger, Beust, Blunt,
  Boccaletti, Bonnefoy, Bonnet, Brandner, Cantalloube, Charnay, Choquet,
  Christiaens, Cl{\'e}net, du~Foresto, Cridland, de~Zeeuw, Dembet, Dexter,
  Drescher, Duvert, Eckart, Eisenhauer, Gao, Garcia, Lopez, Gardner, Gendron,
  Genzel, Gillessen, Girard, Grandjean, Haubois, Hei{\ss}el, Henning, Hinkley,
  Hippler, Horrobin, Houll{\'e}, Hubert, {Jim{\'e}nez-Rosales}, Jocou,
  Kammerer, Kervella, Keppler, Kreidberg, Kulikauskas, Lapeyr{\`e}re, Bouquin,
  L{\'e}na, M{\'e}rand, Maire, Molli{\`e}re, Monnier, Mouillet, M{\"u}ller,
  Nasedkin, Ott, Otten, Paumard, Paladini, Perraut, Perrin, Pueyo, Pfuhl,
  Rameau, Rodet, {Rodr{\'i}guez-Coira}, Rousset, Scheithauer, Shangguan,
  Stadler, Straub, Straubmeier, Sturm, Tacconi, van Dishoeck, Vigan, Vincent,
  von Fellenberg, {Ward-Duong}, Widmann, Wieprecht, Wiezorrek, \&
  Woillez}]{Nowak2020}
Nowak, M., Lacour, S., Lagrange, A.-M., {et~al.} 2020, Astronomy \&
  Astrophysics, 642, L2

\bibitem[{Vousden {et~al.}(2016)Vousden, Farr, \& Mandel}]{Vousden2016}
Vousden, W.~D., Farr, W.~M., \& Mandel, I. 2016, Monthly Notices of the Royal
  Astronomical Society, 455, 1919

\end{thebibliography}

\clearpage

\appendix

\section{Observing logs}
\label{app:logs}

\begin{table}[h!]
  \begin{center}
    \begin{small}
    \begin{tabular}{l l l l l l c c l l}
      \hline
      \hline
      Target & Epoch & Array & Airmass & Seeing & $\tau_0$ &  $\Delta{}\mathrm{RA}$ & $\Delta{}\mathrm{DEC}$ & $\mathrm{NEXP}\times{}$ & Program ID \\
             &       &       & [min-max] & [as] & [ms] & [mas] & [mas] & $\mathrm{NDIT}\times{}\mathrm{DIT}$ & \\
      \hline
      \multirow{6}{*}{HD 73900} & \multirow{2}{*}{2021-12-28} & \multirow{2}{*}{A0-G1-J2-K0} & 1.07-1.08 & 0.40 & 8.51 & $0$ & $0$ & $4\times 16\times 10$~s & \multirow{2}{*}{106.21S5.001} \\
 & & & 1.07-1.09 & 0.41 & 10.38 & $-775$ & $-431$ & $4\times 16\times 10$~s & \\
 & \multirow{2}{*}{2021-10-25} & \multirow{2}{*}{A0-B2-D0-C1} & 1.25-1.30 & 0.52 & 16.49 & $-775$ & $-431$ & $2\times 16\times 10$~s & \multirow{2}{*}{108.22MY.001} \\
 & & & 1.24-1.27 & 0.53 & 16.49 & $0$ & $0$ & $2\times 16\times 10$~s & \\
 & \multirow{2}{*}{2021-11-17} & \multirow{2}{*}{A0-G1-J2-K0} & 1.19-1.22 & 0.71 & 3.25 & $0$ & $0$ & $4\times 16\times 10$~s & \multirow{2}{*}{106.21S5.001} \\
 & & & 1.20-1.24 & 0.81 & 2.80 & $-775$ & $-431$ & $4\times 16\times 10$~s & \\
\hline 
\multirow{4}{*}{HD 25535} & \multirow{2}{*}{2019-10-22} & \multirow{2}{*}{A0-B2-D0-C1} & 1.09-1.17 & 0.61 & 3.24 & $671$ & $870$ & $3\times 32\times 10$~s & \multirow{2}{*}{60.A-9102(H)} \\
 & & & 1.08-1.18 & 0.56 & 3.75 & $0$ & $0$ & $4\times 32\times 10$~s & \\
 & \multirow{2}{*}{2019-10-21} & \multirow{2}{*}{A0-G1-J2-K0} & 1.04-1.11 & 1.34 & 1.59 & $0$ & $0$ & $4\times 32\times 10$~s & \multirow{2}{*}{60.A-9102(H)} \\
 & & & 1.05-1.10 & 1.21 & 1.59 & $671$ & $870$ & $3\times 32\times 10$~s & \\
\hline 
\multirow{4}{*}{HD 91881 } & \multirow{2}{*}{2020-02-20} & \multirow{2}{*}{K0-G2-D0-J3} & 1.44-1.54 & 0.67 & 4.78 & $-1114$ & $724$ & $2\times 8\times 10$~s & \multirow{2}{*}{2104.C-5040(A)} \\
 & & & 1.46-1.52 & 0.64 & 4.65 & $0$ & $0$ & $2\times 8\times 10$~s & \\
 & \multirow{2}{*}{2020-02-06} & \multirow{2}{*}{A0-B2-D0-C1} & 1.11-1.14 & 1.17 & 2.67 & $-1114$ & $724$ & $2\times 16\times 10$~s & \multirow{2}{*}{2104.C-5040(A)} \\
 & & & 1.10-1.12 & 1.00 & 2.58 & $0$ & $0$ & $2\times 16\times 10$~s & \\
\hline 
\multirow{6}{*}{HD 32642} & \multirow{2}{*}{2021-11-03} & \multirow{2}{*}{A0-B2-D0-C1} & 1.76-1.83 & 0.44 & 6.64 & $0$ & $0$ & $2\times 16\times 10$~s & \multirow{2}{*}{108.22MY.001} \\
 & & & 1.79-1.89 & 0.38 & 7.67 & $-872$ & $416$ & $2\times 16\times 10$~s & \\
 & \multirow{2}{*}{2021-11-27} & \multirow{2}{*}{A0-B2-D0-C1} & 1.70-1.76 & 0.57 & 6.35 & $0$ & $0$ & $2\times 16\times 10$~s & \multirow{2}{*}{108.22MY.001} \\
             & & & 1.72-1.81 & 0.57 & 6.35 & $-872$ & $416$ & $2\times 16\times 10$~s & \\
 & \multirow{2}{*}{2021-12-28} & \multirow{2}{*}{A0-G1-J2-K0} & 1.54-1.59 & 0.42 & 8.61 & $-872$ & $416$ & $2\times 16\times 10$~s & \multirow{2}{*}{106.21S5.001} \\
 & & & 1.56-1.60 & 0.41 & 8.61 & $0$ & $0$ & $2\times 16\times 10$~s & \\
\hline
\multirow{2}{*}{HD 1663} & \multirow{2}{*}{2021-10-23} & \multirow{2}{*}{A0-B2-D0-C1} & 1.41-1.43 & 0.48 & 3.34 & $0$ & $0$ & $1\times 8\times 30$~s &  \multirow{2}{*}{108.22MY.001} \\
 & & & 1.36-1.50 & 0.52 & 3.34 & $667$ & $-391$ & $4\times 8\times 30$~s & \\
\hline 
\multirow{4}{*}{HD 123227} & \multirow{2}{*}{2023-03-30} & \multirow{2}{*}{A0-B2-D0-C1} & 1.13-1.14 & 0.85 & 5.59 & $0$ & $0$ & $2\times 16\times 10$~s & \multirow{2}{*}{60.A-9102(J)} \\
 & & & 1.12-1.13 & 0.91 & 5.26 & $-475$ & $905$ & $2\times 16\times 10$~s & \\
 & \multirow{2}{*}{2020-02-07} & \multirow{2}{*}{A0-B2-D0-C1} & 1.39-1.42 & 0.94 & 3.86 & $0$ & $0$ & $2\times 16\times 10$~s & \multirow{2}{*}{2104.C-5040(A)} \\
 & & & 1.40-1.45 & 0.98 & 3.76 & $-535$ & $965$ & $2\times 16\times 10$~s & \\
\hline 
\multirow{12}{*}{$\lambda$ 01 Scl} & \multirow{2}{*}{2021-10-09} & \multirow{2}{*}{A0-G1-J2-K0} & 1.05-1.08 & 0.99 & 1.85 & $0$ & $0$ & $4\times 16\times 10$~s & \multirow{2}{*}{106.21S5.001} \\
 & & & 1.06-1.11 & 0.93 & 2.11 & $337$ & $-425$ & $4\times 16\times 10$~s & \\
 & \multirow{2}{*}{2021-10-22} & \multirow{2}{*}{A0-B2-D0-C1} & 1.04-1.05 & 0.63 & 4.82 & $0$ & $0$ & $2\times 16\times 10$~s & \multirow{2}{*}{108.22MY.001} \\
 & & & 1.04-1.05 & 0.71 & 4.20 & $240$ & $630$ & $2\times 16\times 10$~s & \\
 & \multirow{2}{*}{2021-10-21} & \multirow{2}{*}{A0-B2-D0-C1} & 1.04-1.04 & 0.88 & 2.50 & $240$ & $630$ & $2\times 16\times 10$~s & \multirow{2}{*}{108.22MY.001} \\
 & & & 1.03-1.04 & 0.84 & 2.44 & $0$ & $0$ & $2\times 16\times 10$~s & \\
 & \multirow{2}{*}{2021-10-01} & \multirow{2}{*}{A0-G1-J2-K0} & 1.25-1.30 & 1.19 & 1.81 & $337$ & $-425$ & $2\times 16\times 10$~s & \multirow{2}{*}{106.21S5.001} \\
 & & & 1.28-1.31 & 1.08 & 1.94 & $0$ & $0$ & $2\times 16\times 10$~s & \\
 & \multirow{2}{*}{2021-11-16} & \multirow{2}{*}{A0-G1-J2-K0} & 1.30-1.35 & 0.69 & 3.76 & $240$ & $630$ & $2\times 16\times 10$~s & \multirow{2}{*}{106.21S5.001} \\
 & & & 1.33-1.37 & 0.73 & 3.43 & $0$ & $0$ & $2\times 16\times 10$~s & \\
             & \multirow{2}{*}{2023-10-25} & \multirow{2}{*}{A0-B2-D0-C1} & 1.18-1.35 & 0.76 & 3.59 & $0$ & $0$ & $6\times 16\times 10$~s & \multirow{2}{*}{60.A-9102(J)} \\
 & & & 1.19-1.34 & 0.77 & 3.94 & $-295$ & $-620$ & $4\times 16\times 10$~s & \\      
\hline
\multirow{2}{*}{HD 36584} & \multirow{2}{*}{2021-10-22} & \multirow{2}{*}{A0-G1-J2-K0} & 1.60-1.63 & 0.37 & 7.10 & $549$ & $-1381$ & $2\times 16\times 10$~s & \multirow{2}{*}{106.21S5.001} \\
 & & & 1.62-1.62 & 0.37 & 7.10 & $0$ & $0$ & $1\times 16\times 10$~s & \\
\hline
\multirow{4}{*}{HD 46716} & \multirow{2}{*}{2021-10-22} & \multirow{2}{*}{A0-B2-D0-C1} & 1.23-1.37 & 0.84 & 2.87 & $65$ & $689$ & $4\times 8\times 30$~s & \multirow{2}{*}{108.22MY.001} \\
 & & & 1.22-1.32 & 0.86 & 2.92 & $0$ & $0$ & $4\times 16\times 10$~s & \\
 & \multirow{2}{*}{2021-12-28} & \multirow{2}{*}{A0-G1-J2-K0} & 1.03-1.07 & 0.45 & 6.73 & $65$ & $689$ & $5\times 8\times 30$~s & \multirow{2}{*}{106.21S5.001} \\
 & & & 1.03-1.04 & 0.47 & 6.67 & $0$ & $0$ & $4\times 16\times 10$~s & \\
\hline      
\multirow{2}{*}{HD 46780} & \multirow{2}{*}{2020-12-17} & \multirow{2}{*}{K0-G2-D0-J3} & 1.62-1.63 & 0.44 & 6.54 & $-724$ & $96$ & $2\times 32\times 10$~s & \multirow{2}{*}{106.21S5.001} \\
 & & & 1.62-1.63 & 0.45 & 6.43 & $0$ & $0$ & $2\times 32\times 10$~s & \\
\hline       
\multirow{2}{*}{HD 37904} & \multirow{2}{*}{2020-12-01} & \multirow{2}{*}{A0-G1-J2-K0} & 1.09-1.12 & 0.47 & 4.65 & $-14$ & $-541$ & $4\times 32\times 10$~s & \multirow{2}{*}{106.21S5.001}\\
 & & & 1.11-1.13 & 0.41 & 5.10 & $0$ & $0$ & $4\times 32\times 10$~s & \\
\hline 
\multirow{4}{*}{HD 90444} & \multirow{2}{*}{2022-02-26} & \multirow{2}{*}{A0-G1-J2-K0} & 1.48-1.54 & 0.44 & 13.99 & $0$ & $0$ & $2\times 32\times 10$~s & \multirow{2}{*}{108.22MY.001} \\
 & & & 1.44-1.52 & 0.42 & 14.93 & $411$ & $-688$ & $2\times 32\times 10$~s & \\
 & \multirow{2}{*}{2022-02-28} & \multirow{2}{*}{A0-G1-J2-K0} & 1.62-1.73 & 0.36 & 9.32 & $0$ & $0$ & $2\times 32\times 10$~s & \multirow{2}{*}{106.21S5.001} \\
 & & & 1.56-1.69 & 0.40 & 10.03 & $411$ & $-688$ & $2\times 32\times 10$~s & \\
\hline 
\multirow{2}{*}{HD 174536} & \multirow{2}{*}{2023-03-30} & \multirow{2}{*}{A0-B2-D0-C1} & 1.06-1.15 & 0.74 & 4.98 & $316$ & $-1486$ & $4\times 16\times 10$~s & \multirow{2}{*}{60.A-9102(J)} \\
 & & & 1.06-1.12 & 0.67 & 5.91 & $-0$ & $-0$ & $4\times 16\times 10$~s & \\
\hline 
\multirow{2}{*}{HD 16638} & \multirow{2}{*}{2021-11-17} & \multirow{2}{*}{A0-G1-J2-K0} & 1.88-1.99 & 0.70 & 3.8 & $-266$ & $-574$ & $1\times 16\times 10$~s & \multirow{2}{*}{108.22MY.001} \\
        & & & 1.88-1.99 & 0.69 & 3.9 & $-0$ & $-0$ & $1\times 16\times 10$~s & \\
\hline       
    \end{tabular}
    \caption{Observing log with the Auxiliary Telescopes (ATs).}
    \end{small}    
  \end{center}
\end{table}

\clearpage

\begin{table}
  \begin{center}
    \begin{small}
    \begin{tabular}{l l l l l l c c l l}
      \hline
      \hline
      Target & Epoch & Array & Airmass & Seeing & $\tau_0$ &  $\Delta{}\mathrm{RA}$ & $\Delta{}\mathrm{DEC}$ & $\mathrm{NEXP}\times{}$ & Program ID \\
             &       &       & [min-max] & [as] & [ms] & [mas] & [mas] & $\mathrm{NDIT}\times{}\mathrm{DIT}$ & \\
      \hline
 \multirow{6}{*}{HD 174536} & \multirow{2}{*}{2023-07-02} & \multirow{2}{*}{U1-U2-U3-U4} & 1.12-1.13 & 0.62 & 3.49 & $316$ & $-1486$ & $2\times 32\times 3.0$~s & \multirow{2}{*}{1104.C-0651(G)} \\
 & & & 1.10-1.11 & 0.85 & 2.34 & $-316$ & $1486$ & $2\times 96\times 0.3$~s & \\
 & \multirow{2}{*}{2021-08-27} & \multirow{2}{*}{U1-U2-U3-U4} & 1.06-1.07 & 0.71 & 6.06 & $316$ & $-1486$ & $2\times 12\times 3.0$~s & \multirow{2}{*}{1104.C-0651(D)} \\
 & & & 1.05-1.05 & 0.82 & 4.92 & $-316$ & $1486$ & $2\times 24\times 1.0$~s & \\
 & \multirow{2}{*}{2023-05-07} & \multirow{2}{*}{U1-U2-U3-U4} & 1.02-1.02 & 1.88 & 1.39 & $316$ & $-1486$ & $3\times 32\times 3.0$~s & \multirow{2}{*}{1104.C-0651(G)} \\
 & & & 1.02-1.02 & 2.11 & 1.40 & $-316$ & $1486$ & $2\times 96\times 0.3$~s & \\
\hline 
\multirow{12}{*}{HD 91881} & \multirow{2}{*}{2022-02-20} & \multirow{2}{*}{U1-U2-U3-U4} & 1.03-1.03 & 0.62 & 11.91 & $-1111$ & $700$ & $4\times 96\times 0.3$~s & \multirow{2}{*}{1104.C-0651(B)} \\
 & & & 1.04-1.04 & 0.54 & 11.72 & $1111$ & $-700$ & $3\times 96\times 0.3$~s & \\
 & \multirow{2}{*}{2021-01-07} & \multirow{2}{*}{U1-U2-U3-U4} & 1.06-1.10 & 0.54 & 7.31 & $-1108$ & $710$ & $4\times 64\times 1.0$~s & \multirow{2}{*}{1104.C-0651(C)} \\
 & & & 1.05-1.08 & 0.59 & 7.28 & $1108$ & $710$ & $4\times 64\times 1.0$~s & \\      
 & \multirow{2}{*}{2022-03-21} & \multirow{2}{*}{U1-U2-U3-U4} & 1.04-1.04 & 0.74 & 5.16 & $-1111$ & $700$ & $4\times 48\times 1.0$~s & \multirow{2}{*}{108.226K.002} \\
 & & & 1.05-1.06 & 0.66 & 5.34 & $1111$ & $-700$ & $4\times 48\times 1.0$~s & \\
 & \multirow{2}{*}{2022-01-23} & \multirow{2}{*}{U1-U2-U3-U4} & 1.05-1.06 & 0.28 & 11.67 & $-1108$ & $710$ & $3\times 64\times 1.0$~s & \multirow{2}{*}{1104.C-0651(B)} \\
 & & & 1.04-1.05 & 0.47 & 6.37 & $1108$ & $-710$ & $3\times 64\times 1.0$~s & \\
 & \multirow{2}{*}{2022-03-20} & \multirow{2}{*}{U1-U2-U3-U4} & 1.01-1.32 & 0.96 & 2.35 & $-1111$ & $700$ & $8\times 48\times 1.0$~s & \multirow{2}{*}{108.226K.002} \\
 & & & 1.01-1.25 & 1.28 & 2.43 & $1111$ & $-700$ & $8\times 48\times 1.0$~s & \\
 & \multirow{2}{*}{2023-05-10} & \multirow{2}{*}{U1-U2-U3-U4} & 1.14-1.14 & 0.81 & 7.81 & $-1111$ & $699$ & $2\times 96\times 0.3$~s & \multirow{2}{*}{1104.C-0651(G)} \\
 & & & 1.15-1.16 & 0.72 & 8.32 & $1111$ & $-699$ & $2\times 96\times 0.3$~s & \\
\hline 
\multirow{10}{*}{HD 196885} & \multirow{2}{*}{2021-05-29} & \multirow{2}{*}{U1-U2-U3-U4} & 1.28-1.32 & 0.65 & 4.47 & $313$ & $432$ & $3\times 12\times 30.0$~s & \multirow{2}{*}{5102.C-0122(D)} \\
 & & & 1.25-1.26 & 0.63 & 4.58 & $-313$ & $-432$ & $3\times 48\times 3.0$~s & \\
 & \multirow{2}{*}{2021-08-24} & \multirow{2}{*}{U1-U2-U3-U4} & 1.29-1.33 & 0.83 & 2.37 & $310$ & $425$ & $3\times 12\times 30.0$~s & \multirow{2}{*}{5102.C-0122(D)} \\
 & & & 1.39-1.41 & 1.02 & 2.38 & $-310$ & $-425$ & $2\times 48\times 3.0$~s & \\
 & \multirow{2}{*}{2021-10-20} & \multirow{2}{*}{U1-U2-U3-U4} & 1.48-1.57 & 1.11 & 2.15 & $307$ & $428$ & $3\times 12\times 30.0$~s & \multirow{2}{*}{5102.C-0122(D)} \\
 & & & 1.66-1.73 & 0.90 & 2.55 & $-307$ & $-428$ & $3\times 48\times 3.0$~s & \\
 & \multirow{2}{*}{2021-07-25} & \multirow{2}{*}{U1-U2-U3-U4} & 1.26-1.29 & 0.71 & 3.11 & $309$ & $434$ & $3\times 12\times 30.0$~s & \multirow{2}{*}{5102.C-0122(D)} \\
 & & & 1.32-1.34 & 0.92 & 2.96 & $-309$ & $-434$ & $3\times 48\times 3.0$~s & \\
 & \multirow{2}{*}{2019-08-15} & \multirow{2}{*}{U1-U2-U3-U4} & 1.25-1.26 & 0.96 & 2.46 & $383$ & $429$ & $3\times 20\times 10.0$~s & \multirow{2}{*}{5102.C-0122(B)} \\
 & & & 1.28-1.30 & 0.91 & 2.92 & $-383$ & $-429$ & $3\times 10\times 30.0$~s & \\
\hline 
\multirow{6}{*}{HD 123227} & \multirow{2}{*}{2023-05-10} & \multirow{2}{*}{U1-U2-U3-U4} & 1.19-1.20 & 0.68 & 6.76 & $-401$ & $889$ & $3\times 96\times 0.3$~s & \multirow{2}{*}{1104.C-0651(G)} \\
 & & & 1.17-1.18 & 0.89 & 5.64 & $401$ & $-889$ & $2\times 96\times 0.3$~s & \\
 & \multirow{2}{*}{2023-06-03} & \multirow{2}{*}{U1-U2-U3-U4} & 1.11-1.12 & 0.65 & 7.33 & $-401$ & $889$ & $2\times 96\times 0.3$~s & \multirow{2}{*}{1104.C-0651(G)} \\
 & & & 1.12-1.12 & 0.65 & 4.73 & $401$ & $-889$ & $2\times 96\times 0.3$~s & \\
 & \multirow{2}{*}{2023-05-07} & \multirow{2}{*}{U1-U2-U3-U4} & 1.20-1.21 & 1.51 & 1.77 & $-404$ & $889$ & $3\times 96\times 0.3$~s & \multirow{2}{*}{1104.C-0651(G)} \\
 & & & 1.22-1.22 & 1.26 & 2.06 & $404$ & $-889$ & $2\times 96\times 0.3$~s & \\
\hline 
\multirow{18}{*}{HD 25535} & \multirow{2}{*}{2020-02-11} & \multirow{2}{*}{U1-U2-U3-U4} & 1.05-1.08 & 0.91 & 4.24 & $825$ & $749$ & $4\times 64\times 1.0$~s & \multirow{2}{*}{1103.B-0626(B)} \\
 & & & 1.06-1.10 & 0.77 & 5.97 & $-825$ & $-749$ & $4\times 64\times 1.0$~s & \\
 & \multirow{2}{*}{2023-09-01} & \multirow{2}{*}{U1-U2-U3-U4} & 1.03-1.03 & 1.59 & 1.76 & $726$ & $800$ & $2\times 64\times 1.0$~s & \multirow{2}{*}{111.24MN.001} \\
 & & & 1.02-1.02 & 1.67 & 1.44 & $-726$ & $-800$ & $2\times 64\times 1.0$~s & \\
 & \multirow{2}{*}{2022-08-19} & \multirow{2}{*}{U1-U2-U3-U4} & 1.23-1.24 & 0.48 & 16.64 & $754$ & $786$ & $2\times 64\times 1.0$~s & \multirow{2}{*}{1104.C-0651(F)} \\
 & & & 1.21-1.22 & 0.49 & 14.17 & $-754$ & $-786$ & $2\times 64\times 1.0$~s & \\
 & \multirow{2}{*}{2023-09-25} & \multirow{2}{*}{U1-U2-U3-U4} & 1.05-1.05 & 0.76 & 6.12 & $726$ & $800$ & $2\times 64\times 1.0$~s & \multirow{2}{*}{111.24MN.001} \\
 & & & 1.04-1.04 & 0.86 & 4.81 & $-726$ & $-800$ & $2\times 64\times 1.0$~s & \\
 & \multirow{2}{*}{2022-09-15} & \multirow{2}{*}{U1-U2-U3-U4} & 1.12-1.13 & 0.59 & 3.46 & $752$ & $787$ & $2\times 64\times 1.0$~s & \multirow{2}{*}{1104.C-0651(F)} \\
 & & & 1.11-1.11 & 0.57 & 3.46 & $-752$ & $-787$ & $2\times 64\times 1.0$~s & \\
 & \multirow{2}{*}{2023-11-03} & \multirow{2}{*}{U1-U2-U3-U4} & 1.02-1.02 & 1.11 & 2.97 & $718$ & $804$ & $2\times 64\times 1.0$~s & \multirow{2}{*}{112.25GE.001} \\
 & & & 1.02-1.02 & 1.33 & 2.00 & $-718$ & $-804$ & $2\times 64\times 1.0$~s & \\
 & \multirow{2}{*}{2021-08-28} & \multirow{2}{*}{U1-U2-U3-U4} & 1.25-1.36 & 0.91 & 2.43 & $780$ & $773$ & $4\times 64\times 1.0$~s & \multirow{2}{*}{1104.C-0651(D)} \\
 & & & 1.20-1.31 & 0.97 & 2.85 & $-780$ & $-773$ & $4\times 64\times 1.0$~s & \\
 & \multirow{2}{*}{2019-11-11} & \multirow{2}{*}{U1-U2-U3-U4} & 1.05-1.07 & 0.96 & 2.40 & $825$ & $749$ & $4\times 64\times 1.0$~s & \multirow{2}{*}{1104.C-0651(A)} \\
 & & & 1.04-1.06 & 1.00 & 1.80 & $-825$ & $-749$ & $4\times 64\times 1.0$~s & \\
 & \multirow{2}{*}{2021-08-27} & \multirow{2}{*}{U1-U2-U3-U4} & 1.16-1.18 & 1.42 & 2.38 & $766$ & $785$ & $3\times 64\times 1.0$~s & \multirow{2}{*}{1104.C-0651(D)} \\
 & & & 1.13-1.15 & 1.71 & 1.84 & $-766$ & $-785$ & $3\times 64\times 1.0$~s & \\
\hline 
\multirow{4}{*}{HD 73900} & \multirow{2}{*}{2022-01-25} & \multirow{2}{*}{U1-U2-U3-U4} & 1.06-1.06 & 0.60 & 5.88 & $-827$ & $-456$ & $3\times 96\times 0.3$~s & \multirow{2}{*}{1104.C-0651(B)} \\
 & & & 1.07-1.07 & 0.59 & 6.64 & $827$ & $456$ & $2\times 96\times 0.3$~s & \\
 & \multirow{2}{*}{2022-01-23} & \multirow{2}{*}{U1-U2-U3-U4} & 1.02-1.02 & 0.54 & 15.48 & $-825$ & $-455$ & $3\times 64\times 1.0$~s & \multirow{2}{*}{1104.C-0651(B)} \\
 & & & 1.02-1.02 & 0.42 & 13.22 & $825$ & $455$ & $3\times 64\times 1.0$~s & \\
\hline 
\multirow{4}{*}{HD 36584} & \multirow{2}{*}{2021-08-27} & \multirow{2}{*}{U1-U2-U3-U4} & 1.62-1.64 & 1.44 & 1.80 & $-540$ & $1350$ & $3\times 64\times 1.0$~s & \multirow{2}{*}{1104.C-0651(D)} \\
 & & & 1.59-1.61 & 1.52 & 1.84 & $540$ & $-1350$ & $4\times 64\times 1.0$~s & \\
 & \multirow{2}{*}{2022-09-15} & \multirow{2}{*}{U1-U2-U3-U4} & 1.75-1.77 & 0.60 & 3.62 & $-544$ & $1351$ & $2\times 32\times 3.0$~s & \multirow{2}{*}{1104.C-0651(F)} \\
 & & & 1.71-1.72 & 0.67 & 2.96 & $540$ & $-1350$ & $2\times 32\times 3.0$~s & \\
\hline 
    \end{tabular}
    \caption{Observing log with the Unit Telescopes (UTs).}
    \end{small}    
  \end{center}
\end{table}

\clearpage

\section{Astrometry}
\label{app:astrometry}

\begin{table}[h!]
  \begin{small}
  \begin{center}
    \begin{tabular}{l l l l c c c c c l}
      \hline
      \hline
      \multicolumn{2}{c}{Target} & \multicolumn{2}{c}{Epoch} & $\Delta{}\mathrm{RA}$ & $\Delta{}\mathrm{DEC}$ & $\sigma_{\Delta{}\mathrm{RA}}$ & $\sigma_{\Delta{}\mathrm{DEC}}$ & $\rho$ & Source\\
      WDS identifier & Alt. name & Date & MJD                &  [mas] & [mas] & [mas] & [mas] &  & \\
      \hline
      WDSJ00209+1059 & HD 1663 & 2021-10-23 & 59510.04 & $683.09$ & $-397.77$ & $0.18$ & $0.12$ & $-0.98$ & ATs \\
      \hline      
      \multirow{3}{*}{WDSJ00427-3828} & \multirow{3}{*}{$\lambda$ 01 Scl} & 2021-10-21 & 59508.11 & $-290.45$ & $-624.81$ & $0.05$ & $0.17$ & $-1.00$ & ATs \\
                                 &  & 2021-10-22 & 59509.10 & $-290.79$ & $-624.68$ & $0.09$ & $0.12$ & $1.00$ & ATs \\
                                 &  & 2021-11-16 & 59534.20 & $-290.47$ & $-624.79$ & $0.01$ & $0.09$ & $-1.00$ & ATs \\      
      \hline
      \multirow{11}{*}{WDSJ04021-3429}  & \multirow{11}{*}{HD 25535} & 2019-10-21 & 58777.34 & $826.20$ & $747.81$ & $0.05$ & $0.09$ & $0.84$ & ATs \\
                                 & & 2019-10-22 & 58778.36 & $825.16$ & $748.89$ & $0.09$ & $0.08$ & $0.84$ & ATs \\
                                 & & 2019-11-11 & 58798.18 & $824.75$ & $748.84$ & $0.50$ & $0.50$ & $0.00$ & UTs \\
                                 & & 2020-02-10 & 58890.03 & $818.64$ & $752.26$ & $0.50$ & $0.50$ & $0.00$ & UTs \\
                                 & & 2021-08-26 & 59453.34 & $779.74$ & $773.31$ & $0.50$ & $0.50$ & $0.00$ & UTs \\
                                 & & 2021-08-27 & 59454.31 & $779.51$ & $773.28$ & $0.50$ & $0.50$ & $0.00$ & UTs \\
                                 & & 2022-08-19 & 59810.34 & $754.16$ & $786.14$ & $0.50$ & $0.50$ & $0.00$ & UTs \\
                                 & & 2022-09-15 & 59837.30 & $752.14$ & $787.16$ & $0.50$ & $0.50$ & $0.00$ & UTs \\
                                 & & 2023-08-31 & 60188.39 & $726.79$ & $799.19$ & $0.50$ & $0.50$ & $0.00$ & UTs \\
                                 & & 2023-09-26 & 60212.31 & $728.37$ & $798.47$ & $0.50$ & $0.50$ & $0.00$ & UTs \\
                                 & & 2023-11-02 & 60251.26 & $722.11$ & $801.35$ & $0.50$ & $0.50$ & $0.00$ & UTs \\
      \hline      
      \multirow{3}{*}{WDSJ05055+1948} & \multirow{3}{*}{HD 32642} & 2021-11-03 & 59521.18 & $-861.44$ & $386.17$ & $0.17$ & $0.09$ & $1.00$ & ATs \\
                                 &  & 2021-11-27 & 59545.13 & $-861.94$ & $385.85$ & $0.64$ & $0.06$ & $1.00$ & ATs \\
                                 &  & 2021-12-28 & 59576.20 & $-861.32$ & $386.56$ & $0.01$ & $0.06$ & $1.00$ & ATs \\
      \hline     
      \multirow{3}{*}{WDSJ05270-6837} & \multirow{3}{*}{HD 36584} & 2020-12-01 & 59184.35 & $540.39$ & $-1349.58$ & $0.11$ & $0.10$ & $1.00$ & ATs \\
                                 & & 2021-08-26 & 59453.36 & $543.86$ & $-1350.96$ & $0.50$ & $0.50$ & $0.00$ & UTs \\
                                 & & 2022-09-15 & 59837.28 & $544.79$ & $-1353.06$ & $0.50$ & $0.50$ & $0.00$ & UTs \\
      \hline      
      \multirow{2}{*}{WDSJ06345-1114} & \multirow{2}{*}{HD 46716} & 2021-10-22 & 59509.27 & $116.22$ & $696.42$ & $0.15$ & $0.13$ & $0.87$ & ATs \\
                                 &  & 2021-12-28 & 59576.18 & $116.06$ & $695.74$ & $0.05$ & $0.09$ & $-0.73$ & ATs \\
      \hline
      WDSJ06364+2717 & HD 46780 & 2020-12-17 & 59200.23 & $-713.77$ & $39.29$ & $0.01$ & $0.17$ & $-1.00$ & ATs \\
      \hline      
      \multirow{5}{*}{WDSJ08394-3636} & \multirow{5}{*}{HD 73900} & 2021-10-25 & 59512.34 & $-825.29$ & $-454.65$ & $0.24$ & $0.05$ & $-1.00$ & ATs \\
                                 &  & 2021-11-17 & 59535.29 & $-825.42$ & $-455.38$ & $0.11$ & $0.04$ & $-1.00$ & ATs \\
                                 &  & 2021-12-28 & 59576.22 & $-826.16$ & $-455.88$ & $0.03$ & $0.07$ & $1.00$ & ATs \\
                                 &  & 2022-01-23 & 59602.22 & $-826.61$ & $-456.09$ & $0.50$ & $0.50$ & $0.00$ & UTs \\
                                 &  & 2022-01-24 & 59604.26 & $-826.49$ & $-456.16$ & $0.50$ & $0.50$ & $0.00$ & UTs \\
      \hline      
      \multirow{2}{*}{WDSJ10269+1713} & \multirow{2}{*}{HD 90444} & 2022-02-26 & 59636.26 & $419.24$ & $-715.69$ & $0.04$ & $0.11$ & $1.00$ & ATs \\
                                 &  & 2022-02-28 & 59638.28 & $418.81$ & $-715.40$ & $0.03$ & $0.22$ & $-1.00$ & ATs \\
      \hline      
      \multirow{8}{*}{WDSJ10361-2641} & \multirow{8}{*}{HD 91881} & 2020-02-06 & 58885.17 & $-1105.80$ & $718.54$ & $0.23$ & $0.18$ & $-1.00$ & ATs \\
                                 &  & 2020-02-20 & 58899.36 & $-1106.26$ & $718.39$ & $0.06$ & $1.00$ & $-1.00$ & ATs \\
                                 & & 2021-01-07 & 59221.27 & $-1108.21$ & $710.21$ & $0.50$ & $0.50$ & $0.00$ & UTs \\
                                 & & 2022-01-23 & 59602.24 & $-1110.91$ & $700.41$ & $0.50$ & $0.50$ & $0.00$ & UTs \\
                                 & & 2022-02-19 & 59630.27 & $-1110.77$ & $699.54$ & $0.50$ & $0.50$ & $0.00$ & UTs \\
                                 & & 2022-03-20 & 59658.14 & $-1110.84$ & $698.79$ & $0.50$ & $0.50$ & $0.00$ & UTs \\
                                 & & 2022-03-21 & 59659.20 & $-1110.86$ & $698.83$ & $0.50$ & $0.50$ & $0.00$ & UTs \\
                                 & & 2023-05-10 & 60074.10 & $-1113.06$ & $687.73$ & $0.50$ & $0.50$ & $0.00$ & UTs \\      
      \hline            
      \multirow{5}{*}{WDSJ14077-4952} & \multirow{5}{*}{HD 123227} & 2020-02-07 & 58886.27 & $-475.06$ & $905.84$ & $0.47$ & $0.04$ & $-1.00$ & ATs \\
                                 &  & 2023-03-30 & 60033.30 & $-403.97$ & $889.10$ & $0.08$ & $0.40$ & $1.00$ & ATs \\
                                 & & 2023-05-06 & 60071.24 & $-401.15$ & $888.76$ & $0.50$ & $0.50$ & $0.00$ & UTs \\
                                 & & 2023-05-10 & 60074.08 & $-400.87$ & $888.61$ & $0.50$ & $0.50$ & $0.00$ & UTs \\
                                 & & 2023-06-02 & 60098.11 & $-399.38$ & $888.27$ & $0.50$ & $0.50$ & $0.00$ & UTs \\ 
      \hline
      \multirow{3}{*}{WDSJ18516-1719} & \multirow{3}{*}{HD 174536} & 2021-08-27 & 59454.00 & $-315.59$ & $1486.36$ & $0.50$ & $0.50$ & $0.00$ & UTs \\
                                 & & 2023-03-30 & 60033.39 & $-314.11$ & $1486.51$ & $0.24$ & $0.09$ & $-0.04$ & ATs \\
                                 & & 2023-05-07 & 60071.38 & $314.48$ & $-1486.62$ & $0.50$ & $0.50$ & $0.00$ & UTs \\
                                 & & 2023-07-02 & 60127.13 & $314.35$ & $-1486.42$ & $0.50$ & $0.50$ & $0.00$ & UTs \\
      \hline      
      \multirow{5}{*}{WDSJ20399+1115} & \multirow{5}{*}{HD 196885} & 2019-08-15 & 58710.19 & $377.11$ & $425.59$ & $0.50$ & $0.50$ & $0.00$ & UTs \\
                                 & & 2021-05-29 & 59363.33 & $320.21$ & $428.41$ & $0.50$ & $0.50$ & $0.00$ & UTs \\
                                 & & 2021-07-25 & 59420.26 & $315.09$ & $428.61$ & $0.50$ & $0.50$ & $0.00$ & UTs \\
                                 & & 2021-08-24 & 59450.19 & $312.55$ & $428.32$ & $0.50$ & $0.50$ & $0.00$ & UTs \\
                                 & & 2021-10-20 & 59507.08 & $307.42$ & $428.55$ & $0.50$ & $0.50$ & $0.00$ & UTs \\
      \hline      
    \end{tabular}
    \caption{Astrometric measurements on the vetted binary calibrators.}
    \label{tab:astrometry}
  \end{center}
  \end{small}
\end{table}

\clearpage

\section{Orbital parameters}

\begin{table}[h]
  \begin{small}
  \begin{tabular}{l l l l l l l l l l l l l l l l l l}
    \hline
    \hline
\hline
Parameter & WDSJ00209+1059 & WDSJ00427-3828 & WDSJ04021-3429 & WDSJ05055+1948 & WDSJ05270-6837\\
\hline
\multirow{2}{*}{sma [au]} &129.04 & 85.63 & 67.06 & 129.31 & 75.02\\
 & [106.06-170.56] & [69.83-116.28] & [63.89-73.90] & [110.61-168.25] & [66.35-92.86]\\
\multirow{2}{*}{ecc} &0.51 & 0.42 & 0.41 & 0.42 & 0.79\\
 & [0.29-0.64] & [0.11-0.69] & [0.39-0.44] & [0.28-0.54] & [0.61-0.86]\\
\multirow{2}{*}{inc [deg]} &46.81 & 50.94 & 151.43 & 142.49 & 139.86\\
 & [30.48-57.41] & [28.13-59.93] & [147.42-154.93] & [133.10-157.51] & [123.07-156.19]\\
\multirow{2}{*}{aop [deg]} &172.17 & 184.32 & 233.37 & 173.49 & 192.74\\
 & [42.91-277.20] & [54.72-322.63] & [56.67-239.50] & [77.06-288.48] & [61.59-263.22]\\
\multirow{2}{*}{pan [deg]} &218.13 & 181.86 & 185.47 & 170.28 & 165.36\\
 & [111.50-303.72] & [14.78-207.47] & [22.29-216.86] & [119.00-332.60] & [47.44-312.61]\\
\multirow{2}{*}{tau} &0.73 & 0.36 & 0.10 & 0.61 & 0.67\\
 & [0.63-0.86] & [0.23-0.49] & [0.07-0.13] & [0.50-0.80] & [0.61-0.74]\\
\multirow{2}{*}{plx [mas]} &5.21 & 6.56 & 21.70 & 6.21 & 12.82\\
 & [5.13-5.29] & [6.14-6.97] & [21.64-21.76] & [6.12-6.32] & [12.25-13.40]\\
\multirow{2}{*}{mtot [$M_\odot$]} &5.13 & 5.30 & 2.27 & 3.86 & 1.91\\
 & [4.22-6.14] & [4.29-6.33] & [2.10-2.69] & [3.14-4.87] & [1.51-2.54]\\
\hline
\hline
Parameter & WDSJ06345-1114 & WDSJ06364+2717 & WDSJ08394-3636 & WDSJ10269+1713 & WDSJ10361-2641\\
\hline
\multirow{2}{*}{sma [au]} &83.70 & 33.09 & 26.63 & 38.51 & 37.62\\
 & [74.77-104.50] & [31.94-34.36] & [24.78-28.95] & [33.95-43.54] & [37.11-38.16]\\
\multirow{2}{*}{ecc} &0.40 & 0.75 & 0.85 & 0.95 & 0.75\\
 & [0.31-0.53] & [0.74-0.77] & [0.82-0.88] & [0.94-0.97] & [0.75-0.76]\\
\multirow{2}{*}{inc [deg]} &60.33 & 110.35 & 99.75 & 61.25 & 127.08\\
 & [56.04-64.97] & [109.26-111.38] & [97.63-102.39] & [54.49-65.37] & [126.22-127.94]\\
\multirow{2}{*}{aop [deg]} &183.75 & 275.02 & 145.07 & 180.25 & 218.01\\
 & [30.88-338.50] & [95.57-276.11] & [134.24-321.43] & [7.56-343.66] & [40.79-222.77]\\
\multirow{2}{*}{pan [deg]} &45.38 & 72.25 & 231.70 & 157.85 & 326.13\\
 & [35.58-218.41] & [70.47-251.30] & [52.94-234.80] & [147.77-331.28] & [147.94-329.28]\\
\multirow{2}{*}{tau} &0.74 & 0.17 & 0.49 & 0.62 & 0.54\\
 & [0.61-0.84] & [0.16-0.17] & [0.47-0.51] & [0.61-0.63] & [0.54-0.55]\\
\multirow{2}{*}{plx [mas]} &9.39 & 27.17 & 24.41 & 11.47 & 23.13\\
 & [9.22-9.59] & [26.75-27.59] & [24.37-24.45] & [10.19-13.02] & [23.05-23.21]\\
\multirow{2}{*}{mtot [$M_\odot$]} &3.65 & 2.71 & 4.58 & 3.01 & 2.19\\
 & [2.79-4.88] & [2.42-3.07] & [3.63-5.93] & [2.09-4.43] & [2.12-2.26]\\
\hline
\hline
Parameter & WDSJ14077-4952 & WDSJ18516-1719 & WDSJ20399+1115\\
\hline
\multirow{2}{*}{sma [au]} &64.50 & 1444.18 & 20.73\\
 & [59.32-69.08] & [891.20-2878.30] & [16.84-24.96]\\
\multirow{2}{*}{ecc} &0.12 & 0.53 & 0.42\\
 & [0.08-0.16] & [0.15-0.81] & [0.12-0.68]\\
\multirow{2}{*}{inc [deg]} &59.93 & 63.94 & 123.30\\
 & [57.87-61.25] & [36.50-97.52] & [116.88-138.20]\\
\multirow{2}{*}{aop [deg]} &197.90 & 195.84 & 209.59\\
 & [72.88-297.31] & [73.51-319.94] & [59.93-275.48]\\
\multirow{2}{*}{pan [deg]} &126.68 & 187.47 & 129.32\\
 & [122.73-304.13] & [95.11-338.45] & [68.70-264.59]\\
\multirow{2}{*}{tau} &0.13 & 0.42 & 0.35\\
 & [0.04-0.24] & [0.16-0.78] & [0.22-0.62]\\
\multirow{2}{*}{plx [mas]} &22.23 & 1.19 & 29.41\\
 & [22.15-22.30] & [0.80-1.51] & [29.38-29.44]\\
\multirow{2}{*}{mtot [$M_\odot$]} &2.18 & 2.83 & 1.20\\
 & [2.09-2.27] & [1.38-4.50] & [0.88-1.99]\\
\hline
  \end{tabular}
  \end{small}   
  \caption{Orbital parameters for our 13 UT binary calibrators.\\
  Notes: For each parameter, we give the best estimate, which corresponds to the mode of the posterior distribution, as well as the $68\%$ confidence interval reported by orbitize!. The parameter tau corresponds to the time at periapsis, expressed in unit of orbital period from the reference epoch $58849~\mathrm{MJD}$.}
  \label{tab:parameters}
\end{table}

\end{document}